\documentclass{elsart}

\usepackage{graphicx}
\usepackage{amssymb}

\usepackage{url}

\newcommand\noff{n_{\rm off}}     % y        total observed in background region
\newcommand\non{n_{\rm on}}       % x        total observed in signal region
\newcommand\ntot{n_{\rm tot}}     % k        total observed in both regions

\newcommand\muon{\mu_{\rm on}}    %          mu_s + mu_b  = total mean in signal region
\newcommand\muoff{\mu_{\rm off}}  % 
\newcommand\mutot{\mu_{\rm tot}}  %          mu total in signal plus background regions

\newcommand\binp{\rho}            %          binomial parameter for muoff/(muon+muoff)
\newcommand\ratmean{\lambda}      %          ratio of poisson means muon/muoff

\newcommand\lhood{{\cal L}}
\newcommand\deltalhood{\Delta(-2\ln\lhood)}
\newcommand\bi{{\rm Bi}}
\newcommand\poi{{\rm Poi}}

\newcommand\tlow{t_{\rm low}}
\newcommand\tmax{t_{\rm max}}
\newcommand\tup{t_{\rm up}}
\newcommand\tmin{t_{\rm min}}

\newcommand\zbi{Z_{\rm Bi}}

\newcommand\epsdir{.}

\begin{document}

\begin{frontmatter}



\title{Frequentist Evaluation of Intervals Estimated for
a Binomial Parameter and for the Ratio of Poisson Means}


\author{Robert D. Cousins},
\ead{cousins@physics.ucla.edu}
\author{Kathryn E. Hymes},
\author{Jordan Tucker}
\ead{tucker@physics.ucla.edu}
\address{Dept.\ of Physics and Astronomy, University of 
California, Los Angeles, California 90095, USA}
 
\begin{abstract}
Confidence intervals for a binomial parameter or for the ratio of
Poisson means are commonly desired in high energy physics (HEP)
applications such as measuring a detection efficiency
or branching ratio.  Due to the discreteness of
the data, in both of these problems the
frequentist coverage probability unfortunately depends on the unknown
parameter.  Trade-offs among desiderata have
led to numerous sets of intervals in the statistics literature,
while in HEP one typically encounters only the classic
intervals of Clopper-Pearson (central intervals with no
undercoverage but substantial over-coverage) or a few approximate
methods which perform rather
poorly.  If strict coverage is relaxed, some sort of averaging is
needed to compare intervals. In most of the statistics
literature, this averaging
is over different values of the unknown parameter, which is
conceptually problematic from the frequentist point of view in which
the unknown parameter is typically fixed.  In contrast, we perform an
(unconditional) {\it average over observed data} in the
ratio-of-Poisson-means problem.  If strict conditional coverage is desired,
we recommend
Clopper-Pearson intervals and intervals from inverting the likelihood
ratio test (for central and 
non-central
intervals, respectively).  Lancaster's mid-$P$ modification to
either provides excellent unconditional average coverage
in the ratio-of-Poisson-means problem.
\end{abstract}

\end{frontmatter}

% main text
\section{Introduction}
\label{intro}
The construction of confidence intervals for a binomial parameter
(probability of success in a binomial distribution), while already
performed by Clopper and Pearson (C-P) in 1934 \cite{clopper34},
remains a topic of discussion in the modern statistics literature due
to differences in opinion about the best way to deal with imperfect
coverage rooted in the discreteness of the observed number of
successes.  Clopper and Pearson's central intervals, while
guaranteeing no undercoverage, result in considerable overcoverage
(conservatism). Numerous alternatives have been put forward in the
intervening years, with reviews such as that by Brown, Cai, and
Dasgupta \cite{brown01} recommending for general use some sets of
intervals which are less conservative than those of C-P, but which
undercover for certain values of the binomial parameter.  In this
paper, we examine the problem from the point of view of high energy
physics (HEP) applications, including the problem of confidence
intervals for the ratio of Poisson means.  The latter problem provides
an additional frequentist criterion, not yet considered by Brown et
al., for judging the merits of sets of intervals for a binomial
parameter.

\begin{figure}
  \centering
  \includegraphics*[width=0.5\textwidth]{\epsdir/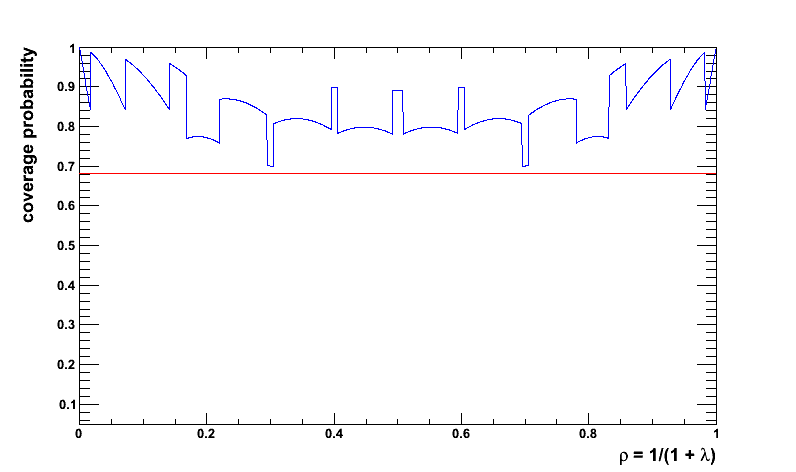}
  \put(-50,20){\bf\large (a)}
  \includegraphics*[width=0.5\textwidth]{\epsdir/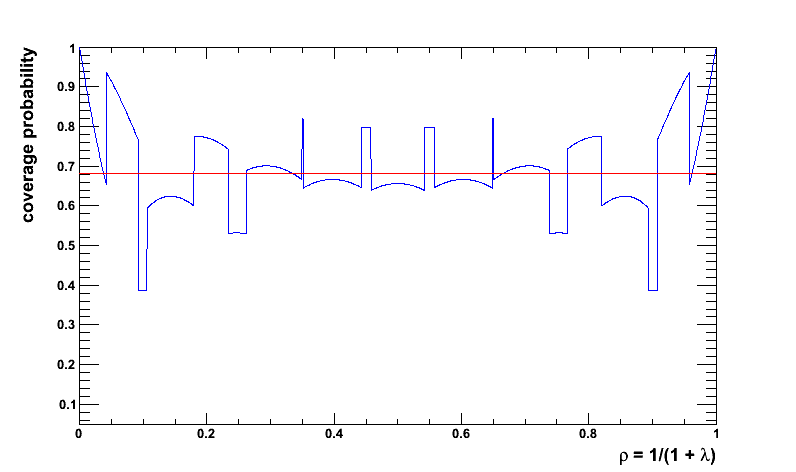}
  \put(-50,20){\bf\large (b)}
  \caption{
(a) Coverage of 68.27\% C.L. Clopper-Pearson intervals, and (b)
coverage of intervals calculated using a Bayesian method with Jeffreys
prior and containing 68.27\% posterior probability, both as a function
of $\binp$, for fixed $\ntot=10$. (a) and (b) are horizontal slices of
Figs.~\ref{fig:cpjef2}a and b, respectively.
}
  \label{fig:cpjefrho}
\end{figure}

Figure~\ref{fig:cpjefrho}a illustrates the issue to be addressed.  For
each value of the binomial parameter $\binp$, one supposes that it is
the true but unknown value, and calculates the long-run fraction of
experiments for which that value is contained in (``covered by'') the
reported confidence intervals.  In Fig.~\ref{fig:cpjefrho}a, the number of
trials is fixed at 10, the probabilities for the number of successes
are calculated from the binomial formula using the true value of
$\binp$, and the central C-P confidence intervals with a confidence
level (C.L.)  of 68.27\% are used. The upper and lower endpoints of
the C-P interval are, respectively, 15.87\% C.L. lower and upper
one-sided confidence limits.  The coverage of the one-sided confidence
limits is always greater than or equal to 15.87\%, with equality on a
discrete finite set of values.  As this set of points is different for
upper and lower confidence limits, the coverage of the two-sided
intervals in this example is always strictly greater than 68.27\%, an
unfortunate consequence of the discrete nature of the observation.

For comparison, Fig.~\ref{fig:cpjefrho}b is the coverage plot for central
intervals derived using a Bayesian technique with Jeffreys prior, as
described below.  The coverage oscillates around the nominal 68.27\%,
in a way that by eye seems to have an ``average'' value near 68.27\%.
The problem from the frequentist point of view is that such averaging
over values of the unknown parameter is typically not appropriate
since the unknown true value of $\binp$ is fixed, i.e., not sampled
from a distribution.

\begin{figure}
  \centering
  \includegraphics*[width=0.5\textwidth]{\epsdir/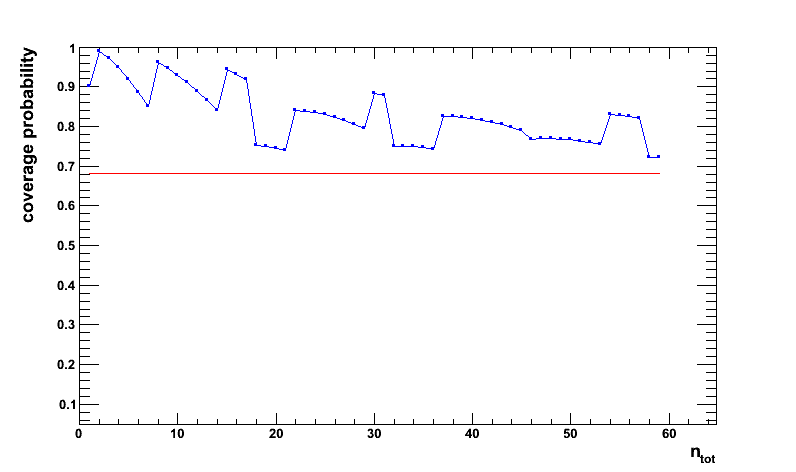}
  \put(-50,20){\bf\large (a)}
  \includegraphics*[width=0.5\textwidth]{\epsdir/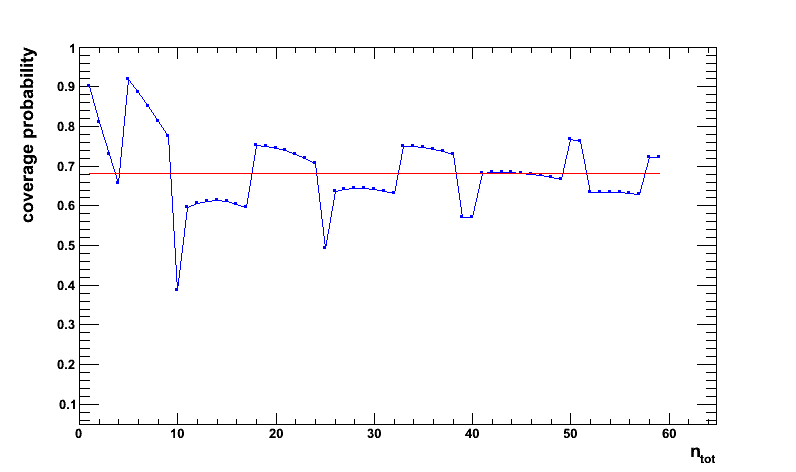}
  \put(-50,20){\bf\large (b)}
  \caption{
(a) Coverage of 68.27\% C.L. Clopper-Pearson intervals, and (b)
coverage of intervals calculated using a Bayesian method with Jeffreys
prior and containing 68.27\% posterior probability, as a function of
$\ntot$, for fixed $\binp=0.1$. (a) and (b) are vertical slices of
Figs.~\ref{fig:cpjef2}a and b, respectively.
}
  \label{fig:cpjefn}
\end{figure}

The effect of discreteness is also displayed in Figs.~\ref{fig:cpjefn}a
and b, which show the coverage as a function of $\ntot$,
for fixed $\binp=0.1$.  The above four plots are horizontal and
vertical slices of a much larger pattern of behavior displayed in
Figs~\ref{fig:cpjef2}a and b.  In these two figures, and
corresponding figures below, $\Delta {\rm CL}$ is the difference
between the actual coverage and the nominal coverage, in this case
68.27\%.

\begin{figure}
  \centering
  \includegraphics*[width=0.5\textwidth]{\epsdir/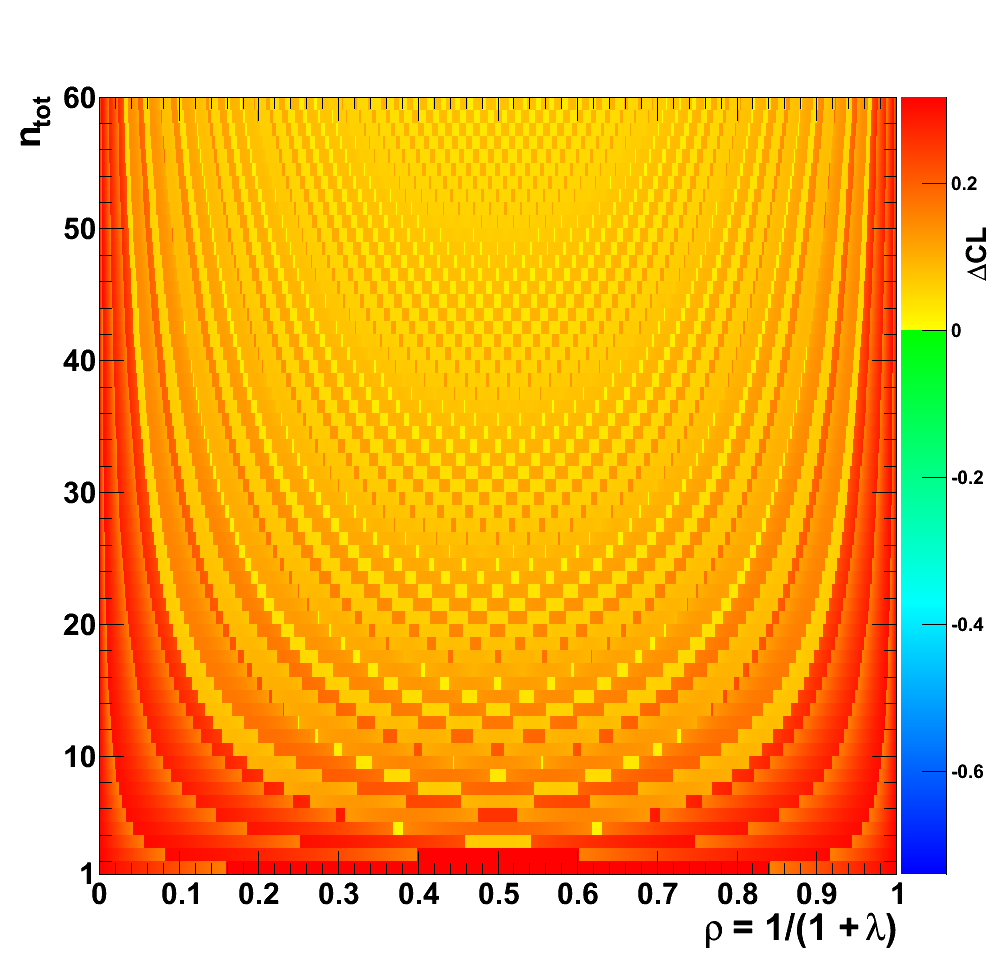}
  \put(-175,0){\bf\large (a)}
  \includegraphics*[width=0.5\textwidth]{\epsdir/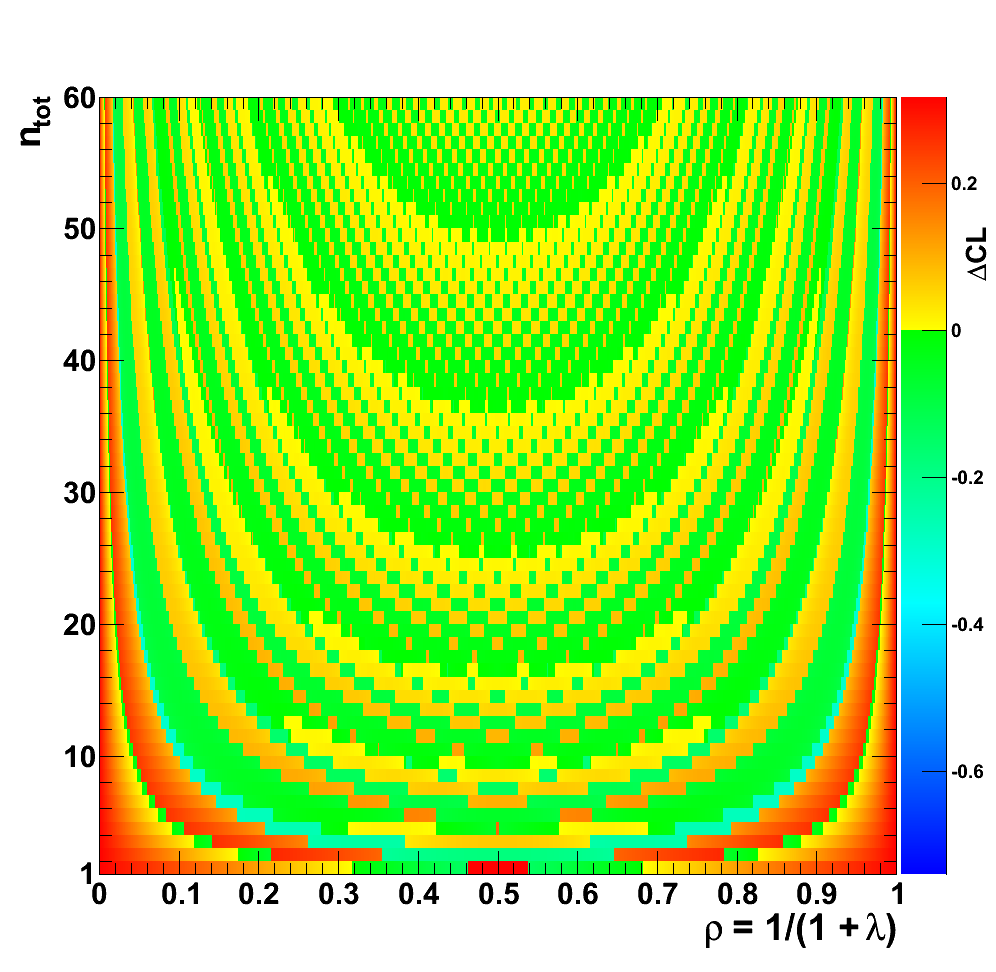}
  \put(-175,0){\bf\large (b)}
  \caption{
(a) Coverage of 68.27\% C.L. Clopper-Pearson intervals, and (b)
coverage of intervals calculated using a Bayesian method with Jeffreys
prior and containing 68.27\% posterior probability, as a function of
$\binp$ and $\ntot$.  $\Delta {\rm CL}$ is the difference between the
actual coverage and the nominal coverage, 68.27\%.
}
  \label{fig:cpjef2}
\end{figure}

These are but two of many sets of intervals that have been proposed.
The saw-tooth features of the coverage plots are intrinsic to all
methods except the randomization technique (mentioned in
Sec.~\ref{def}) which brings other difficulties.  Which sets are
deemed preferable depends on the value one attaches to never having
undercoverage, on what sort of averaging (if any) over values of
$\binp$ one allows, whether or not one desires central intervals, and
additional issues such as whether one is especially concerned about
behavior near the endpoints, $\binp = 0$ and 1.

In this paper, we emphasize that a {\it frequentist} averaging method,
which averages over repeatedly sampled {\it data}, can be used to
evaluate sets of intervals, in contrast to most previous averaging
methods which average over the parameter $\binp$ in some metric. The
frequentist average is performed by using the strong connection
between confidence intervals for a binomial parameter and confidence
intervals for the ratio of two unknown Poisson means.  For pairs of
integers sampled from two fixed but unknown Poisson means,
fluctuations in the total number of observed events provides a random
sampling which partially smoothes out the saw-tooth structure seen in
binomial coverage plots.  Said another way (using terminology defined
below) we use the unconditional global coverage as a criterion for
averaging over imperfect conditional coverage of each fixed total
number of events.

In the traditional definition of ``confidence interval'', defined by
Neyman as we discuss below, the name implies no undercoverage for any
value of the unknown parameter.  When dealing with approximate
methods, immaterial departures from perfect coverage are typically
tolerated as long as it is clearly understood that coverage is only
approximate.  When methods yield intervals which are known to have
non-negligible undercoverage for some values of the unknown parameter
(such as for the mid-$P$ intervals for the binomial parameter), the
statistics literature is mixed on whether or not to refer to these
intervals as confidence intervals.  In this paper, we attempt to
follow HEP practice by requiring no undercoverage when referring to
intervals as ``confidence intervals''.

In Sec.~\ref{def}, we review the relevant concepts from interval and
hypothesis test construction and define the notation.  In
Sec.~\ref{binom}, we briefly describe a number of papers from the vast
literature on binomial intervals.  In Sec.~\ref{ratio}, we generalize
to the ratio-of-Poisson-means problem, and review some relevant
literature.  In Sec.~\ref{performance} we present our results on the
coverage of a number of the methods.  We conclude in
Sec.~\ref{conclusion}.

\section{Definitions and Notation}
\label{def}
We let $\bi(\non|\ntot,\binp)$ denote the probability of $\non$
successes in $\ntot$ trials, each with binomial parameter $\binp$:
\begin{equation}
\label{binomial}
\bi(\non|\ntot,\binp) = \frac{\ntot!}{\non!(\ntot-\non)!}\, 
\binp^{\non}\,(1-\binp)^{(\ntot-\non)}.
\end{equation}
In repeated trials, $\non$ has mean 
\begin{equation}
\ntot\binp
\end{equation}
and rms deviation
\begin{equation}
\label{eqn:rms}
\sqrt{\ntot\binp(1-\binp)}.
\end{equation}
For asymptotically large $\ntot$, Bi can be approximated by a normal
distribution with this mean and rms deviation.

With observed number of successes $\non$, the likelihood function
$\lhood(\binp)$ follows from reading Eqn.~\ref{binomial} as a function
of $\binp$.  The maximum is at
\begin{equation}
\hat\binp = \non/\ntot.
\end{equation}

In some applications, $\ntot$ is not fixed but is itself a random
variable sampled from a Poisson distribution with mean $\mutot$:
\begin{equation}
\poi(\ntot | \mutot) =  \frac{\e^{-(\mutot)}\,
(\mutot)^{\ntot}}{\ntot !}.
\end{equation}
In this case, $\non$ and $\noff = \ntot - \non$ can be considered to
be independent random variables, each sampled from a Poisson
distribution with means $\muon$ and $\muoff$, respectively, satisfying
\begin{equation}
\muon+\muoff=\mutot.
\end{equation} 
The ratio of the Poisson means is then
\begin{equation}
\ratmean=\muoff/\muon, 
\end{equation} 
and the binomial parameter can be written as
\begin{equation}
\label{binrat}
\binp = \muon/\mutot = 1/(1+\ratmean).
\end{equation} 
The joint probability $P(\non,\noff)$ of observing $\non$ and $\noff$
can then be expressed in two equivalent ways: as the product of
independent Poisson probabilities for $\non$ and $\noff$; or as the
product of a single Poisson probability with mean $\mutot$ for the
total number of events $\ntot$, and the binomial probability that this
total is divided as observed:
\begin{eqnarray}
P(\non,\noff) & = & 
 \frac{\e^{-\muon} \muon^{\non}}{\non !} \times
 \frac{\e^{-\muoff} \muoff^{\noff}}{\noff !} \nonumber \\ & = &
 \frac{\e^{-(\muon+\muoff)}\,
(\muon+\muoff)^{\ntot}}{\ntot !}  \times \ \\
&& \frac{\ntot!}{\non!(\ntot-\non)!}\, 
\binp^{\non}\,(1-\binp)^{(\ntot-\non)}.
\label{eqn-jointProb}
\end{eqnarray}
In more compact notation, we have:
\begin{eqnarray}
P(\non,\noff) 
&=& \poi(\non| \muon)\ \poi(\noff| \muoff) \\
&=&\poi(\ntot| \mutot)\ \bi(\non| \ntot, \binp).
\label{poibi}
\end{eqnarray}
This observation is the basis of hypothesis tests on the ratio of
Poisson means going back to Przyborowski and Wilenski \cite{Przy} in
1940, as recommended in HEP by James and Roos \cite{jamesroos}, and as
discussed by statistician Reid~\cite{Reid95}.  All the dependence on
ratio of Poisson means $\ratmean$ is in the {\em conditional} binomial
probability for the observed ``successes'' $\non$, {\em given} the
observed total number of events $\ntot = \non+\noff$.

We consider a general parameter $\theta$ (such as $\binp$ or
$\ratmean$) and randomly sampled data (such as $\non$ or other
observables), the probability of which depends on $\theta$.  We then
consider a recipe for computing the endpoints of a confidence interval
$[\tlow,\tup]$ for $\theta$, as functions of the (randomly sampled)
data.  (In this paper we always include the endpoints in the
confidence interval.)  The set of all confidence intervals obtainable
from all possible data sets using this recipe is called a {\it
confidence set}.  For each value of $\theta$, one can then compute the
probability that that $\theta$ is contained in (``covered by'') the
confidence intervals in the confidence set, for data sampled according
to that $\theta$.  Normally it is highly desirable that this coverage
probability be independent of $\theta$, and is called the
confidence coefficient or (more commonly in HEP) the confidence level
(C.L.) of the confidence set.  For situations such as those in this
paper, in which the data takes on only discrete values, the coverage
probability depends on $\theta$, as illustrated above in
Figs.~\ref{fig:cpjefrho}a and b.

In classical hypothesis testing, a common hypothesis test is that
which tests the hypothesis $H_0$ that $\theta$ is equal to a
particular value, $\theta_0$, against the alternative that
$\theta\ne\theta_0$.  One constructs recipes for accepting/rejecting
$H_0$ based on the (randomly sampled) data, the probability for which
depends on $\theta$.  One defines the significance level $\alpha$ of
the test (also called size of the test) as the probability of
rejecting $H_0$ if is true; again it is desirable that $\alpha$ is
independent of $\theta$.  In the formal theory of Neyman-Pearson
hypothesis testing, $\alpha$ is specified in advance; once data are
obtained, the {\em $p$-value} is the smallest value of $\alpha$ for
which $H_0$ would be rejected.

As discussed by Kendall and Stuart and successors \cite{kendall}, one
can construct a hypothesis test at significance level $\alpha$ simply
by using a confidence set with C.L. $ = 1-\alpha$ and accepting $H_0$
if $\theta_0$ is contained in the confidence interval for $\theta$
based on the obtained data. One can equally well derive confidence
sets from any given recipe for testing the hypothesis
$\theta=\theta_0$, simply by including in the interval those values of
$\theta_0$ which would not be rejected by such a test.  This way of
constructing confidence intervals is referred to in the statistics
literature as ``inverting the hypothesis test''.  (An example now
familiar in the HEP literature is the set of intervals advocated by
Feldman and Cousins \cite{feldmancousins}, which are constructed by
inverting the likelihood ratio test of Ref.~\cite{kendall}.)  It can
happen that the resulting ``intervals'' are not simply connected, in
which case various adjustments are typically made, for example adding
to the interval any interior points not initially part of it (thus
adding to the over-coverage).

In this duality, confidence intervals formed by inverting a test with
significance level $\alpha$ have coverage probability $= 1-\alpha$
under $H_0$, i.e.,
\begin{equation}
\label{cover}
P(\theta_0 \in [\tlow,\tup]) = 1-\alpha.
\end{equation}
{\em Central} confidence intervals have the additional property that
the intervals $[\tlow,\tmax]$ and $[\tmin,\tup]$ each separately have
coverage probability $ = 1 - \alpha/2$, i.e.,
\begin{equation}
\label{central}
P(\theta_0 \in [\tlow,\tmax]) = P(\theta_0 \in [\tmin,\tup]) = 1-\alpha/2,
\end{equation}
where $\tmax$ and $\tmin$ are the maximum and minimum values of
$\theta$ defined in the model (e.g., $\tmin=0$ and $\tmax=1$ if
$\theta$ is a binomial parameter).  In this case, for example, $\tup$
is often referred to as a $(1-\alpha/2)$ C.L. upper confidence limit
for $\theta$.

If, due to the discreteness, the significance level can only be
specified to be less than or equal to $\alpha$, then the equal signs
in Eqns.~\ref{cover} and \ref{central} become ``$\ge$''.

Without invoking a Gaussian approximation in the construction of an
interval itself, it is often useful to make the correspondence with
the number of Gaussian standard deviations having a {\it
single}-tailed probability equal to $\alpha/2$.  Thus, $Z$ denotes the
number of standard deviations away from the center of a Gaussian
distribution, with a subscript representing the (one-tailed) tail
probability beyond that $Z$.
\begin{equation}
\label{zdef}
Z_{\alpha/2} = \Phi^{-1}(1-\alpha/2) = -\Phi^{-1}(\alpha/2)
\end{equation}
where
\begin{equation}
\label{Phidef}
\Phi(Z) = \frac{1}{\sqrt{2\pi}} \int_{-\infty}^Z \,\exp(-t^2/2)\,dt
\ =\ \frac{1 + {\rm erf}(Z/\sqrt{2})}{2},
\end{equation}
so that
\begin{equation}
\label{eqn:z}
Z = \sqrt{2}\, {\rm erf}^{-1}(1-\alpha).
\end{equation}
E.g., $Z_{\alpha/2} = 1$ for $\alpha/2 =
0.159$, and $Z_{\alpha/2} = 1.64$ for $\alpha/2 = 0.05$.

\section{Recipes for intervals for $\binp$}
\label{binom}

A plethora of recipes exists for intervals approximating confidence
intervals for binomial parameter $\binp$. They correspond to various
choices regarding:
\begin{itemize}
\item Whether or not the intervals are central intervals;
\item Whether or not the intervals are derived from rigorously
inverting a hypothesis test (in which case, which test?);
\item Whether or not an asymptotic approximation is invoked;
\item Whether or not Bayesian machinery is used to derive the
intervals;
\item Whether or not so-called ``corrections'' are made in an attempt
to improve the coverage probability.
\end{itemize}
As emphasized by Cai \cite{cai05}, some methods with bad properties as
one-sided intervals have good properties as two-sided intervals due to
cancellations in coverage departures between the two tails.

\subsection{Asymptotic approximations}

We begin with one of the most popular methods, which is also one of
the worst-performing if not {\em the} worst-performing of popular
methods.  We follow the literature in referring to this interval as
the {\it Wald interval}.  After estimating
\begin{equation}
\hat\binp = \non/\ntot,
\end{equation}
the Wald method invokes the Gaussian approximation {\it without
properly inverting the hypothesis test against the null}, but rather
simply substituting $\hat\binp$ for $\binp$ into Eqn.~\ref{eqn:rms}
and using this {\it fixed} value of the rms to obtain the two endpoints,
\begin{equation}
\label{wald}
\hat\binp \pm Z_{\alpha/2} \sqrt{ \frac{\hat\binp(1-\hat\binp)}{\ntot} }.
\end{equation}

Already in 1927, Edwin Wilson \cite{wilson27} realized that since the
rms depends on the unknown parameter $\binp$, the more appropriate way
to invoke the Gaussian approximation was by consistently inverting the
test using the rms of the null hypothesis for each value of $\binp$.
For the lower endpoint, one uses the lowest value $\binp_1$ such that
$\binp_1 + Z_{\alpha/2}\sqrt{\binp_1(1-\binp_1)/\ntot}$ contains
$\hat\binp$.  Analogously for the upper endpoint, one uses the largest
value $\binp_2$ such that $\binp_2 -
Z_{\alpha/2}\sqrt{\binp_2(1-\binp_2)/\ntot}$ contains $\hat\binp$.
Letting $T = (Z_{\alpha/2})^2/\ntot$, this leads to a quadratic
equation in $\binp$ for the endpoints, $(\binp - \hat\binp)^2 =
T\binp(1-\binp)$, with solutions
\begin{equation}
\label{wilson}
\binp = \frac{\hat\binp + T/2}{1+T} \pm 
\frac{\sqrt{\hat\binp(1-\hat\binp)T + T^2/4}}{1+T}.
\end{equation}
These endpoints form the {\it Wilson score interval}; in spite of the
fact that it is a non-iterative solution using nothing more than a
square root, sadly it is commonly overlooked in favor of the Wald
interval when a quick Gaussian estimate is desired.

Letting $\tilde\binp$ denote the midpoint of the Wilson score interval,
from Eqn.~\ref{wilson} one has
\begin{equation} 
\tilde\binp = \frac{\hat\binp + T/2}{1+T} =
\frac{\non + (Z_{\alpha/2})^2/2}{\ntot + (Z_{\alpha/2})^2}.
\end{equation}
As discussed in detail by Agresti and Coull \cite{agresticoull98},
$\tilde\binp$ differs from $\hat\binp$ by formally adding
$(Z_{\alpha/2})^2$ to the number of actual trials, and making half of
them successes.  It thus ``shrinks'' the maximum-likelihood point
estimate $\hat\binp$ towards 0.5.  For 95\% C.L., ($Z_{\alpha/2} =
1.96$), the easy-to-remember rule of thumb is simply ``add four trials
with two successes'' to obtain the (approximate) Wilson midpoint.  For
quick estimates one can use $\tilde\binp$ rather than $\hat\binp$ (and
$\ntot + (Z_{\alpha/2})^2$ rather than $\ntot$) in the Wald formula
(Eqn. \ref{wald}) and obtain intervals with surprisingly decent
coverage, much better than when using $\hat\binp$ (and avoiding the
useless result at extreme data).  We refer to such intervals as {\it
generalized Agresti-Coull (AC) intervals} (adding ``generalized'' to
the name given by Brown et al. \cite{brown01} to distinguish from the
simpler version). Agresti and Coull themselves (who regard the C-P
intervals as not optimal for statistical practice due to their
conservatism) advocate AC intervals for teaching and the Wilson score
interval for statistical practice \cite{agresticoullcomment}.

The asymptotic theory in which log likelihood-ratios are related to
chi-square distributions \cite{wilks38,jamescpc} provides another
interval estimate.  In the present case, the interval consists of all
points satisfying
%e.g. vollset 93, newcombe98
\begin{equation}
\label{wilks}
Z_{\alpha/2}^2\ \ge\ -2\ln\frac{\lhood(\binp)}{\lhood(\hat\binp)} 
\  =\  2\ln\left(\frac{\hat\binp}{\binp}\right)^{\non} + 
       2\ln\left(\frac{1-\hat\binp}{1-\binp}\right)^{(\ntot-\non)}.
\end{equation}
As there is more than one use of the likelihood ratio for intervals in
this paper, we refer unambiguously to intervals from Eqn.~\ref{wilks}
as {\it $\deltalhood$ intervals}.  In addition to the usual caution
required in using asymptotic formulas for small numbers of events, in
the present case there are well-known issues at the extrema of
$\binp$, where the conditions of the asymptotic theory justifying
Eqn.~\ref{wilks} are not satisfied.

As discussed by Cox and Hinkley \cite{coxhinkley74}, for the
exponential family of distributions, i.e., those of the form
$p(\theta) = \exp(a(\theta)b(x) + c(\theta) + d(x))$, the
transformation to the ``natural parameter'' $\phi = -a(\theta)$ and
new data variable $z=b(x)$ leads to some mathematical simplifications.
The natural parameter for the binomial distribution is the logit
function,
\begin{equation}
\phi = \ln(\binp/(1-\binp)), 
\end{equation}
also known as the log odds ratio; it is a convenient map from (0,1) to
$(-\infty,\infty)$ in a variety of contexts. Such non-linear maps are
a reminder that the concept of ``shortest'' is metric-dependent: it is
easy to find pairs of intervals whose relative length in $\binp$ is
reversed when transformed to $\phi$.

The logit makes the mathematics simpler, but as Cox and Hinkley note,
whether this is really the best parametrization can depend on other
considerations as well. Models involving the logit and its inverse
have a long history and were used in the work that was awarded the
2000 Nobel Memorial Prize in Economics.  In any case, one can apply
the same sort of Gaussian approximation to the logit $\phi$ as is
applied in forming the Wald intervals for $\binp$.  The maximum
likelihood estimate $\hat\phi$ is obtained by plugging in
$\hat\binp$. The variance of $\hat\phi$ is estimated as
$\ntot/(\non(\ntot - \non)) = 1/(\hat\binp(1-\hat\binp))$.  One then
has an interval for $\phi$, which can be mapped into an interval for
$\binp$.  As the formulas are singular for $\non=0$ and $\non=\ntot$,
patches are required, which are sometimes used for other values of
$\non$ as well, in particular adding 1/2 to both numerator and
denominator in the logit formula \cite{brown01,gart66,pricebonett00}.

\subsection{Neyman's Construction: ``Exact'' Inversion of a Hypothesis 
Test}

Given any test statistic and an ordering defined for it, confidence
intervals with minimum guaranteed coverage can be constructed by the
technique of Neyman \cite{neyman37}, which corresponds to inverting a
hypothesis test with rigorously calculated significance level.  Such
methods are often called ``exact'' since approximations are not made
in the calculation of the probabilities, but as already shown for
Clopper-Pearson, the coverage is by no means ``exactly'' equal to the
nominal C.L.!  Analogous to the Neyman construction described in
detail for a similar discrete problem in Ref.~\cite{feldmancousins},
for each value of $\binp$ one forms acceptance intervals by adding the
probabilities $\bi(\non|\ntot,\binp)$ for observed $\non$ until the
threshold $1-\alpha$ is crossed. An auxiliary principle for the
ordering in which the probabilities for the $\non$ are to be added to
the acceptance set must be specified.

Clopper and Pearson \cite{clopper34} constructed central confidence
intervals which remain the standard \cite{pdg} for those who insist (as has been
common in HEP) that coverage is always rigorously respected; the
ordering is performed separately on each end of the acceptance
interval.  As noted above, the cost is severe over-coverage for some
values of $\binp$.  Angus and Schafer \cite{angusschafer84} compute
over-coverage of C-P intervals, pointing out that $(1-\alpha)$
C.L. intervals can have coverage probability as high as $(1-\alpha/2)$
for some values of the true $\binp$; in fact the coverage is always
this high or larger if $\ntot$ is small enough that $\ntot < (1 -
\ln\alpha/\ln 2)$.

Sterne \cite{sterne54}, followed soon by Crow \cite{crow56},
constructed sets of {\it non-central} intervals with guaranteed
minimum coverage.  The idea is to reduce over-coverage due to
discreteness by relaxing the requirement in Eqn.~\ref{central} while
retaining that in Eqn.~\ref{cover}.  An obvious ordering principle to
start with is based on $\bi(\non|\ntot,\binp)$, i.e., adding points to
(either end of) the acceptance interval in decreasing order of
probability so as to minimize the length of the acceptance interval.
There is room for adjustment, however, since in many cases the
acceptance interval can be shifted, keeping its length fixed while
still maintaining coverage.  As there is considerable ambiguity in the
best way to make such adjustments, there have been numerous attempts
to improve upon Sterne's non-central intervals, variously referred to
as {\it two-tailed} or {\it both-tailed} intervals.

Blyth and Still \cite{blythstill83} give a very detailed discussion of
the ambiguities encountered in such both-tailed constructions. They
list some desirable features of intervals and, while giving their
preferences for resolving ambiguities, note that ``We see no way of
combining these desirable properties into a precise criterion that
would be generally accepted.''  Casella \cite{casella86} reviews Blyth
and Still and their predecessors and describes a method for
systematically further reducing the length of confidence intervals
obtained from such constructions: ``\dots move all the lower endpoints
of the intervals as far to the right as possible.''  In commenting
\cite{casellacomment} on Brown et al., he strongly advocates covering
at the nominal value or greater, preferring the Blyth-Still intervals
with his length-reduction algorithm.  Blaker
\cite{blaker00,blakerspjotvoll00} discusses in enlightening detail
various both-tailed methods, arriving at intervals which have good
properties (nesting) when viewed as a function of the confidence
level.  But Vos and Hudson \cite{voshudson08} explain in detail how
both-tailed tests, even those of Blaker \cite{blaker00}, inevitably
have some undesirable behavior due to discreteness.

In HEP, Feldman and Cousins \cite{feldmancousins} popularized a Neyman
construction that is equivalent to inverting the hypothesis test based
on likelihood ratios.  The likelihood-ratio ordering in the Neyman
construction is based not on $\bi(\non|\ntot,\binp)$ as used by
Sterne, but on the likelihood ratio
$\bi(\non|\ntot,\binp)/\bi(\non|\ntot,\hat\binp)$. The corresponding
test, the likelihood ratio test (LR test), is one of the standard
methods in classical statistics \cite{kendall}. Coverage of
both-tailed intervals for $\binp$ from such ``exact inversion of the LR
test'' was illustrated by statisticians Corcoran and Mehta
\cite{corcoranmehtacomment}, who prefer over-coverage to
under-coverage, and who advocate either these intervals or the
Blyth-Still-Casella intervals.  Ranucci \cite{ranucci} compares
coverage plots of intervals for $\binp$ from such likelihood-ratio
ordering with the intervals of C-P and of Sterne.  As mentioned above
(and described in Sec. IV of Ref.~\cite{feldmancousins}), some
interior points can be absent from the ``interval'' after first
inverting the LR test; if so, in the present paper they are added in
order to make the interval simply connected.

\subsubsection{Randomization, Mid-$P$, and Continuity Correction}

In order to remove the over-coverage in Neyman constructions caused by
the discreteness of the integer-valued observations such as that of
C-P, in 1950 Stevens \cite{stevens50} and others suggested adding a
random number uniform on (0,1) to the observed integer, and performing
the construction on the resultant continuous variable.  As discussed
in detail in Ref.~\cite{kendall}, this technique, known as {\em
randomization}, results in shorter intervals and perfect coverage. But
as this extra random number was to be chosen from a table of (uniform)
random numbers, it is rarely if ever used except in theoretical
discussions.  Reference~\cite{cousins94} discusses how more meaningful
data-based uniform variates can be justifiably used in randomization
of Poisson observations, but we do not pursue this approach in this
paper.

As an alternative to randomization, Lancaster \cite{lancaster61}
suggested in 1961 to deal with the discreteness issue in many
distributions by quoting an intermediate value of the tail
probability, since known as the ``mid-$P$'' value.  For a one-sided
test, it is the null probability of more extreme results {\it plus
(only) half the probability of the observed $\non$}.  It corresponds
to randomization always with the addition of 1/2 to the observed
integer successes rather than addition of a uniform variate on (0,1).
By using only half the probability rather than all the probability of
the observed $\non$, the mid-$P$ is less than the strict $p$-value.
As such, it has neither perfect coverage nor a guarantee against
undercoverage, but the mid-$P$ has attracted much more of a following
than randomization, as the result is not influenced by an arbitrary
random number.  Berry and Armitage \cite{berryarmitage95} review
mid-$P$ intervals in various contexts including the
binomial problem, suggesting that they can be appropriate when
combining results from several studies.  Agresti and Gottard
\cite{agrestigottard05,agrestigottard07} further advocate mid-$P$
intervals, provide a useful overview, and provide a
function for computing them in the statistical package $R$.

Another commonly used device in dealing with discrete distributions is
called (somewhat optimistically) a {\it continuity correction}, for
example adding or subtracting 1/2 (or more generally another constant)
from the observed number of successes.  Although there is some
advocacy of continuity corrections with respect to the binomial
problem in the literature, it appears that there are better-performing
ways to deal with the discreteness \cite{brown01,brown05}.

\subsection{Bayesian-inspired methods}

Intervals derived using Bayesian machinery \cite{ohagan} can be
evaluated according to their frequentist coverage properties, and
there has long been interest in prior probability density functions
(``priors'') which lead to Bayesian credible intervals possessing
approximate frequentist coverage.  Recent reviews of such
``probability matching priors'' are in
Refs.~\cite{reid03,bainesmeng07}.  Since the work of Welch and Peers
\cite{welchpeers63,welch65,peers65}, it has been recognized that
Jeffreys's prior \cite{jeffreys61,kass96} (derived by Jeffreys under a
different motivation) is the lowest-order probability matching prior
for one parameter (although care must be taken in interpreting this
result for a discrete distribution such as binomial). The Jeffreys
prior for the binomial problem is
\begin{equation}
p(\binp) \propto \frac{1}{\sqrt{\binp(1-\binp)}},
\end{equation}
which is a special case (with $a=b=1/2$) of the two-parameter {\it
beta distribution}, which has pdf
\begin{equation}
p(\binp\,;a,b) \propto 
\binp^{a-1} (1-\binp)^{b-1}.
\end{equation}
The beta distribution is closely linked to the binomial distribution
\cite{ohagan}, and varying $a$ and $b$ provides a family of priors
(including the uniform prior with $a=b=1$).  The posterior from a beta
prior is also a beta distribution \cite{ohagan,berger85}, and
intervals can be obtained from it using various criteria such as
length or centrality.

Geisser \cite{geisser84} considers several noninformative priors in
the Bayesian literature and advocates a prior uniform in $\binp$,
rejecting the Jeffreys prior because it violates the (strong)
likelihood principle \cite{ohagan}.  The Comments following
Ref.~\cite{geisser84} (by Bernardo, Novick, and Zellner) point out
problems when $\binp$ is near 0 or 1.  Brenner and Quan
\cite{brennerquan90} also advocate a prior uniform in $\binp$,
apparently unaware of the many issues \cite{kass96,geisser84} in
trying to represent ``no prior information'' in a prior.  Copas
\cite{copas92} emphasizes that Bayesian-derived results, such as those
of Brenner and Quan, do not automatically have good frequentist
properties, and in particular criticize the prior uniform in $\binp$.

Rubin and Schenker \cite{rubin87} derive logit-based intervals using
the Jeffreys prior, recalling earlier work (including Gart
\cite{gart66}) connecting this approach to using asymptotic logit
estimation after appending a half success and a half failure as
mentioned above.  They calculate coverage both for fixed values of
$\binp$ and for values averaged over the Jeffreys prior.

\subsection{Comparative studies}
Given the abundance of methods, a number of authors have compared them
by various criteria such as average coverage or average length (both
of which are metric dependent), behavior near the extrema of $\binp$,
etc.  There is no general agreement on even basic features, such as
whether or not coverage should be respected everywhere or in an
average sense.  And as noted above, preferences can differ if one is
concerned only with one-sided intervals.

Reiczigel \cite{reiczigel03} advocates quoting an adjusted
significance level based on calculated coverage rather than the
nominal coverage used in the construction.  Agresti \cite{agresti03}
advocates inverting two-tailed tests (leading to non-central
intervals) rather than two one-tailed tests.  Puza and O'Neill
\cite{puza06} perform coverage studies and advocate a ``new class'' of
C-P-inspired intervals which transition from one-sided to two-sided
intervals.

Vollset \cite{vollset93} reviews in detail thirteen methods,
recommending a continuity-corrected Wilson score method (strongly
disfavored by Refs.~\cite{brown01,brown05}), but describing as
``safe'' the C-P intervals, mid-$P$ intervals, and Wilson score
intervals without the continuity correction.  He finds
likelihood-ratio intervals to be too narrow for boundary outcomes.

Edwardes \cite{edwardes98} compared several methods using coverage
averaged over a chosen metric, studying the behavior as a function of
the constant used in the continuity correction. Among many results, he
finds good performance for a Wald logit interval with {\it negative}
continuity correction.

Newcombe \cite{newcombe98} considers the ``strict conservatism'' of
the C-P method to be ``unnecessarily conservative'' and compares it to
the Wald and score methods, with and without continuity corrections,
mid-$P$, and $\deltalhood$ methods, and appears to favor the mid-$P$
and score methods.

Lu \cite{wangshulu00} compares lengths of intervals made with Bayesian
methods with beta function priors, with endpoints adjusted according
to Blyth \cite{blyth86}. He discusses in some detail the
beta-distribution formulas and their numerical evaluation.

Agresti and Min \cite{agrestimin01} generally prefer both-tailed
(non-central) tests if coverage is strictly required (unless one
specifically requires a one-sided-test), and recommend mid-$P$ tests
if not.  They also discuss using unconditional coverage in eliminating
nuisance parameters in the context of the difference in binomial
parameters.

Pires and Amado \cite{pires08} compare 20 methods (counting various
continuity corrections), with a table giving formulas for all of them.
They prefer C-P intervals if coverage is strictly required, or the
(ungeneralized) Agresti-Coull ``add 4'' method \cite{agresticoull98}
if not.

Brown et al. \cite{brown01} consider the Clopper-Pearson intervals to
be ``wastefully conservative'', and advocate three sets of intervals
with coverage oscillating about the nominal value: the Wilson score
interval, the Agresti-Coull interval, and a Bayesian interval with
Jeffreys prior and equal tails (except when $\non$ is at the extreme
values, in which case they take one tail).

Coverage plots are illustrated in Refs.~\cite{brown01,
corcoranmehtacomment, ranucci,agrestigottard07, copas92, rubin87,
reiczigel03, agresti03, puza06, vollset93, edwardes98, agrestimin01,
pires08}.

\section{Application to the Ratio of Poisson Means}
\label{ratio}

As discussed in Sec.~\ref{def}, intervals for the ratio of Poisson
means $\ratmean$ are readily obtained from intervals for the binomial
parameter $\binp$, and vice versa.  The {\it conditional coverage of
$\ratmean$, given $\ntot$}, can be read off coverage plots for $\binp$
using $\ratmean = (1-\binp)/\binp$.  However, we can also consider the
{\it unconditional coverage of $\ratmean$}, as a function of the two
unknown means $\muon$ and $\muoff$, as follows.

\begin{figure}
  \centering
  \includegraphics*[width=0.5\textwidth]{\epsdir/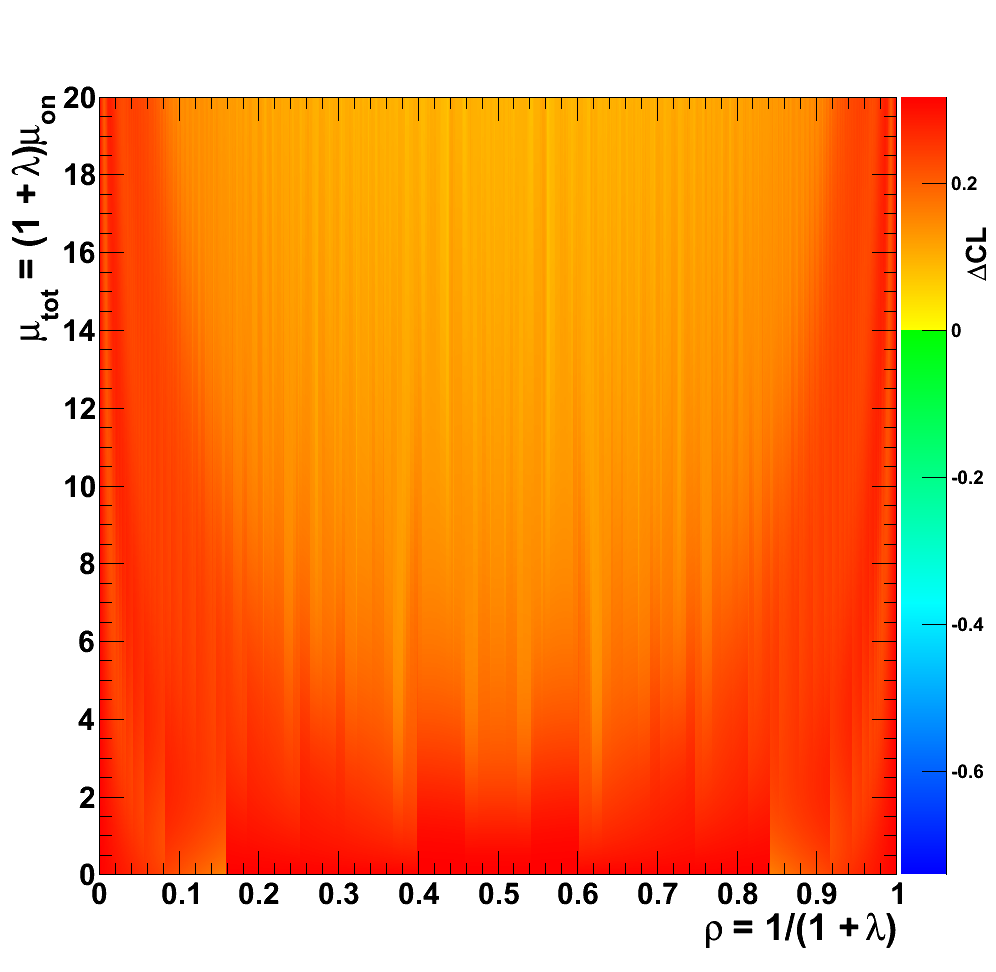}
  \put(-175,0){\bf\large (a)}
  \includegraphics*[width=0.5\textwidth]{\epsdir/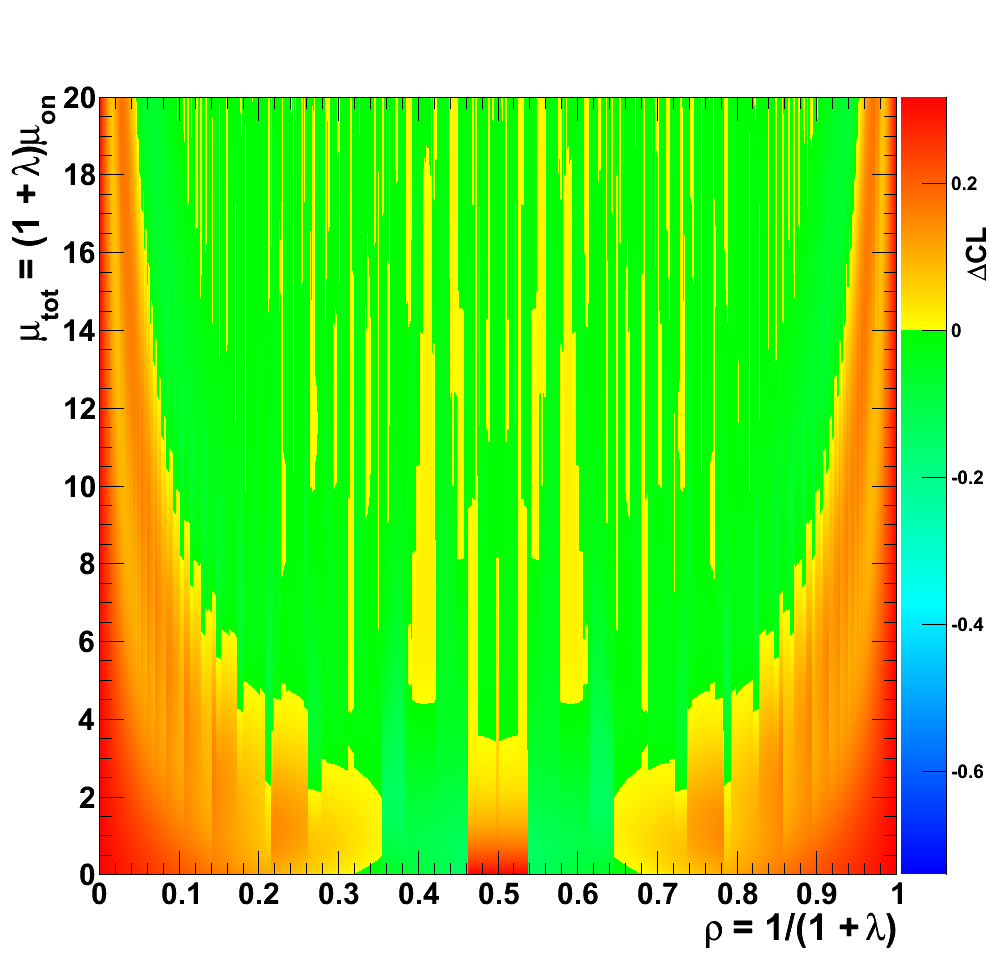}
  \put(-175,0){\bf\large (b)}
  \caption{
Unconditional coverage of (a) 68.27\% C.L. Clopper-Pearson intervals
for the ratio of Poisson means $\ratmean$ and (b) intervals for
$\ratmean$ calculated using a Bayesian method with Jeffreys prior and
containing 68.27\% posterior probability.  As described in the text,
the coverage as a function of $(\muon,\muoff)$ is displayed
equivalently as a function of $(\binp,\mutot)$.  $\Delta {\rm CL}$ is
the difference between the actual coverage and the nominal coverage,
68.27\%.
}
  \label{fig:cpjefrat}
\end{figure}

\begin{figure}
  \centering
  \includegraphics*[width=0.5\textwidth]{\epsdir/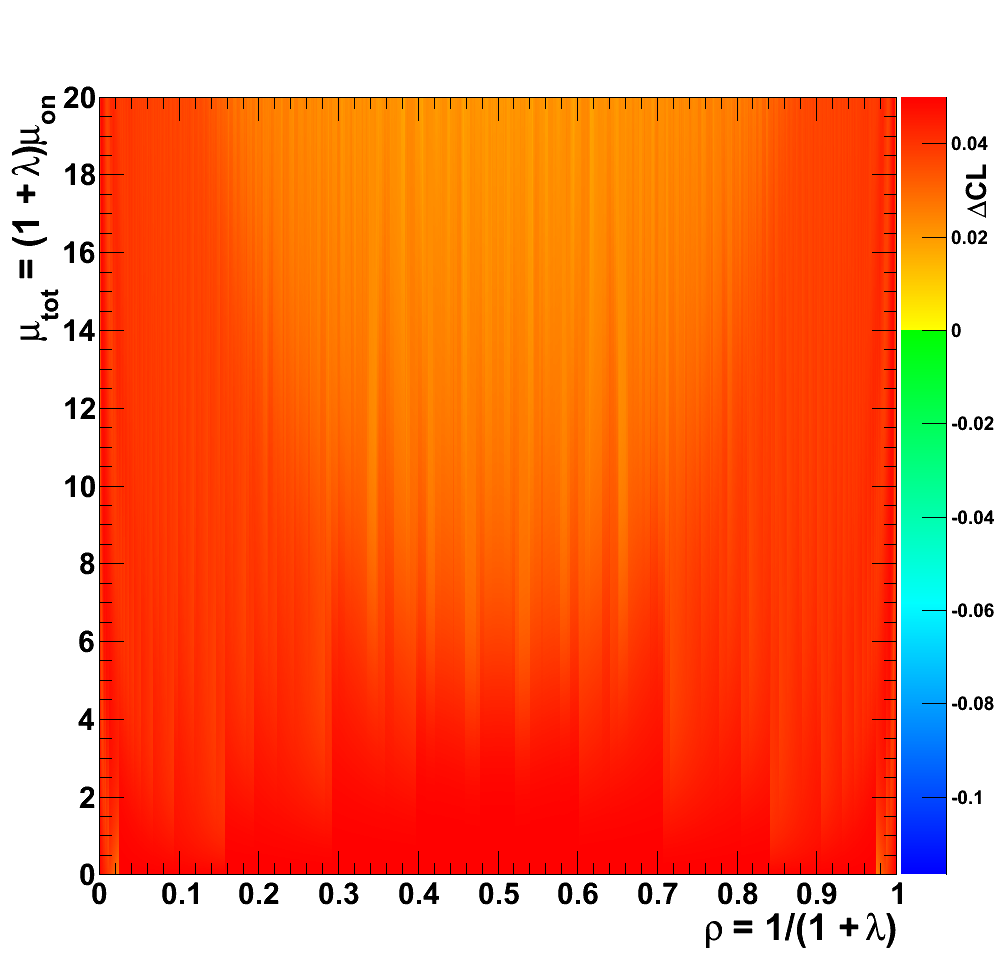}
  \put(-175,0){\bf\large (a)}
  \includegraphics*[width=0.5\textwidth]{\epsdir/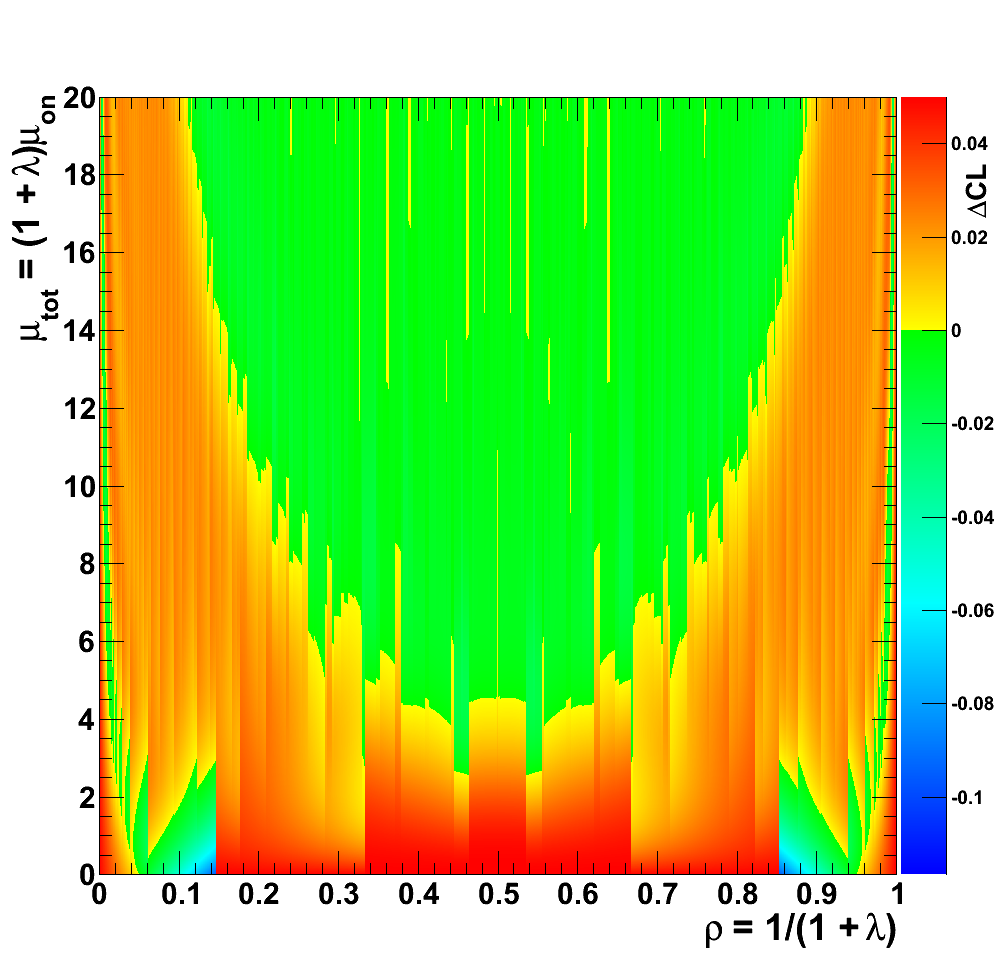}
  \put(-175,0){\bf\large (b)}
  \caption{
Unconditional coverage of (a) 95\% C.L. Clopper-Pearson intervals for
$\ratmean$ and (b) intervals for $\ratmean$ calculated using a
Bayesian method with Jeffreys prior and containing 95\% posterior
probability.
}
  \label{fig:cpjefrat95}
\end{figure}

Given fixed $\muon$, $\muoff$ (and hence $\ratmean$), a C.L., and a
recipe for intervals, then for all pairs $(\non,\noff)$, one can
calculate both the confidence interval for $\lambda$ ($[\tlow,\tup]$)
and the probability $P(\non,\noff)$ of obtaining that pair
(Eqn.~\ref{eqn-jointProb}).  From these one can calculate
probabilities that $\ratmean<\tlow$, that $\ratmean \in [\tlow,\tup]$
and that $\ratmean>\tup$.  Figures~\ref{fig:cpjefrat}a
and b have the results of such unconditional coverage
for the Clopper-Pearson and Jeffreys-prior-based recipes used in the
previous figures; Figs.~\ref{fig:cpjefrat95}a and b
contain the corresponding plots for 95\% C.L.  In order to facilitate comparison with the
conditional coverage plots, the axes are $\mutot$ and $\binp$, from
which one can make the translation to $(\muon,\muoff)$ via $\muon =
\binp\mutot$ and $\muoff = (1-\binp)\mutot$.

These plots of unconditional coverage thus {\it average over observed
$\ntot$} given true values of $\muon$ and $\muoff$, in contrast to
nearly all previous studies which average over unknown true values of
parameters.  While the use of the unconditional ensemble (rather than
the restricted ``conditional'' ensemble having the observed $\ntot$)
goes against the mainstream statistical practice of using the
conditional ensemble in a case such as this \cite{Reid95}, we believe
that this frequentist averaging over data provides at least as good a
way to average-out some discreteness effects as does the common
averaging over $\binp$, which requires a choice of metric (often
$\binp$ itself, although one can argue that the metric in which the
prior is uniform is the natural metric in a Bayesian calculation).
The issue is discussed in detail in Ref.~\cite{cousinsratio}, which as
mentioned below describes a construction of central confidence
intervals for the ratio of Poisson means having strict unconditional,
but not conditional, coverage. (Averaging over observed {\it data}
with different values of $\ntot$ was used in Ref.~\cite{cousinsratio}
to cancel out some under- and over-coverage in different values of
$\ntot$ at each value of the ratio.)

As apparent from Fig.~\ref{fig:cpjefrat}a, applying Clopper-Pearson
binomial confidence intervals to the ratio-of-Poisson-means problem
further propagates the over-coverage due to the discreteness.  We
return to this important point in Sec.\ref{performance} below after
first briefly reviewing some previous work applying non-C-P binomial
intervals to the ratio-of-Poisson-means problem.

Price and Bonett \cite{pricebonett00} consider various solutions to
the problem of the ratio of Poisson means from a broad point of view,
including translating into this problem the binomial confidence
intervals of C-P \cite{clopper34}, Wilson score \cite{wilson27}, and
Agresti and Coull \cite{agresticoull98}.  They also consider recipes
derived directly for the ratio problem, namely a square-root
transformation, an adjusted Wald log-linear model equivalent to the
adjusted Wald logit formula mentioned above, and Bayesian methods
including a Gamma prior for the ratio.  Their conclusions depend as
usual on considerations such as whether coverage is rigorously
required, but tend to favor the adjusted Wald log-linear model in
which 0.5 is added to the observed counts, resulting in endpoints
\begin{equation}
\frac{\non + 0.5}{\ntot-\non + 0.5}\ \exp\left( \pm Z_{\alpha/2} 
\sqrt{\frac{1}{\non+0.5} + \frac{1}{\ntot-\non+0.5}} \right)
\end{equation}

Tang and Ng \cite{tang04}, in commenting on a paper by Graham et
al. \cite{graham03}, examine several methods for intervals for the
ratio of Poisson means, including several based on binomial methods.
They prefer the adjusted Wald logit method also favored by Price and
Bonett, citing them as the source.

Barker and Caldwell \cite{barkercadwell08} compare results of eight
methods for 95\% C.L. intervals, including Bayesian with uniform and
Jeffreys prior.  They prefer the Wald log linear method
% (As they describe it, ``constructed using the asymptotic
% normality of (\ln X - \ln Y - \ln \theta)/(1/X + 1/Y)^{0.5}.''
but do not mention the adjustment of adding 0.5 (nor do they cite
Price and Bonett); they use instead the C-P interval when
$\min(\non,\ntot-\non)=0$.  They found that this composite set
maintains coverage (even though its theoretical justification relies
on asymptotic approximation) and generally performs better than other
methods which maintain coverage. If some under-coverage is allowed,
they favor Bayesian with uniform prior and the Wilson score interval.
(Their criteria include length in the metric uniform in $\binp$.)

Gu et al. \cite{gu08} compare four general approaches via Monte Carlo
simulation, restricting themselves to the one-sided test.  They prefer
a test based on a variance-stabilizing transformation of Huffman
\cite{huffman84}, using an idea of Anscombe \cite{anscombe48}. A
likelihood ratio test is most powerful against the alternatives they
considered, but as it can undercover they advise caution in its use.

Cousins \cite{cousinsratio} describes his multi-dimensional Neyman
construction to obtain a set of {\it central} confidence intervals for
the ratio of Poisson means, obtaining intervals that are strictly
conservative in unconditional coverage, and which are always subsets
(proper subsets except for $\ntot=1$) of Clopper-Pearson-derived
intervals.  The unconditional coverage is obtained by averaging over
conditional under-coverage and over-coverage at different values of
$\ntot$. When translated back into confidence intervals for a binomial
parameter, these intervals are remarkably similar to
mid-$P$ intervals, as discussed below.

In performing coverage tests, there is an issue of what to do if the
data obtained has {\it both} $\noff$ and $\non$ equal to zero, so that
$\ntot=0$.  As nothing has been learned about the ratio, the only
sensible confidence interval is $(0,\infty)$, which always covers the
unknown true value.  Cousins argues in Ref.~\cite{cousinsratio} that
such experiments should be excluded from the coverage calculation,
since as a practical matter, ``An experimenter who observes neither
Poisson process will normally make {\em no} statement regarding the
ratio of their means!  Thus, practical confidence intervals should
have the property that the requisite coverage is obtained when one
considers only those experimenters who do not obtain (0,0).''  We
still believe this to be the case, but note that the coverage
calculations of
Refs.~\cite{pricebonett00,tang04,barkercadwell08}
include observed data (0,0).

In the remainder of this paper, we discuss in more detail how the
ratio-of-Poisson-means problem allows one to evaluate binomial
intervals using a frequentist average over data; we find this to be
preferable to averaging over the binomial parameter, which requires a
choice of metric (or a choice of Bayesian prior from which a
corresponding natural metric can be inferred).  We then perform this
evaluation for a number of the available sets of intervals, and
conclude with observations and recommendations.

\section{Frequentist evaluation of the performance of the various 
recipes}
\label{performance}

Given $\ntot$, $\binp$, and a C.L., one can use any of the above
recipes to obtain the set of $\ntot+1$ intervals
$[\tlow,\tup]$ for $\binp$ corresponding to the possible observations.
As described and illustrated above, in the frequentist evaluation of
these intervals, one calculates the probability
$\bi(\non|\ntot,\binp)$ of obtaining each interval, and
thus the probabilities that $\binp<\tlow$, that $\binp \in
[\tlow,\tup]$ and that $\binp>\tup$. For the ratio of Poisson means
problem, one is given $(\non,\noff)$, and a C.L., from which
$\ntot=\non+\noff$ and hence an interval for $\rho$ is calculated
and then translated into an interval for $\ratmean$ using
Eqn.~\ref{binrat}.  For any given $\mutot$ and $\ratmean$,
probabilities of obtaining $(\non,\noff)$ are calculated from
Eqn.~\ref{poibi}, and thus unconditional probabilities for covering
$\ratmean$ can be calculated from the obtained interval sets and
these probabilities.

\begin{figure}
  \centering
  \includegraphics*[width=0.5\textwidth]{\epsdir/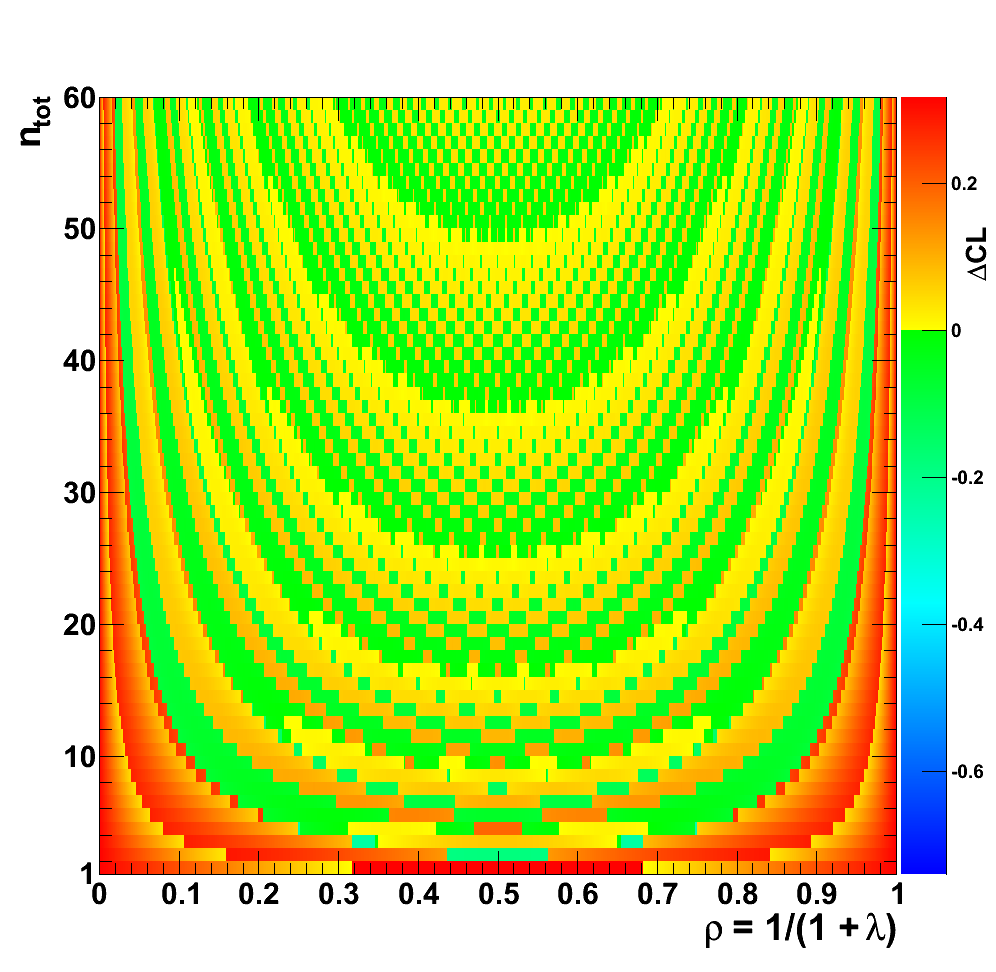}
  \put(-175,0){\bf\large (a)}
  \includegraphics*[width=0.5\textwidth]{\epsdir/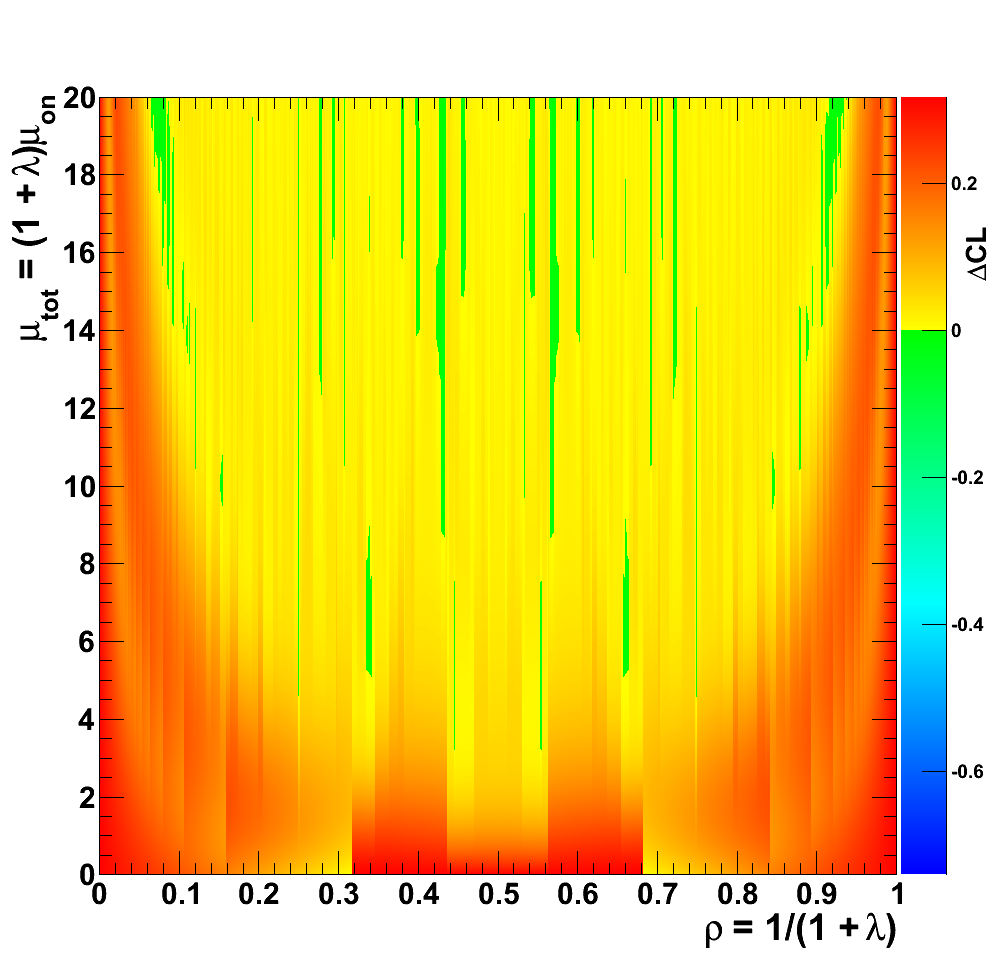}
  \put(-175,0){\bf\large (b)}
  \caption{
(a) Coverage of 68.27\% C.L. (Clopper-Pearson) mid-$P$ intervals, as a
function of $\binp$ and $\ntot$, and (b) unconditional coverage of the
same intervals for $\ratmean$. A horizontal slice of (b) is in
Fig.~\ref{fig:cpmidpcousinsrho}a.
}
  \label{fig:cpmidp2rat}
\end{figure}

\begin{figure}
  \centering
  \includegraphics*[width=0.5\textwidth]{\epsdir/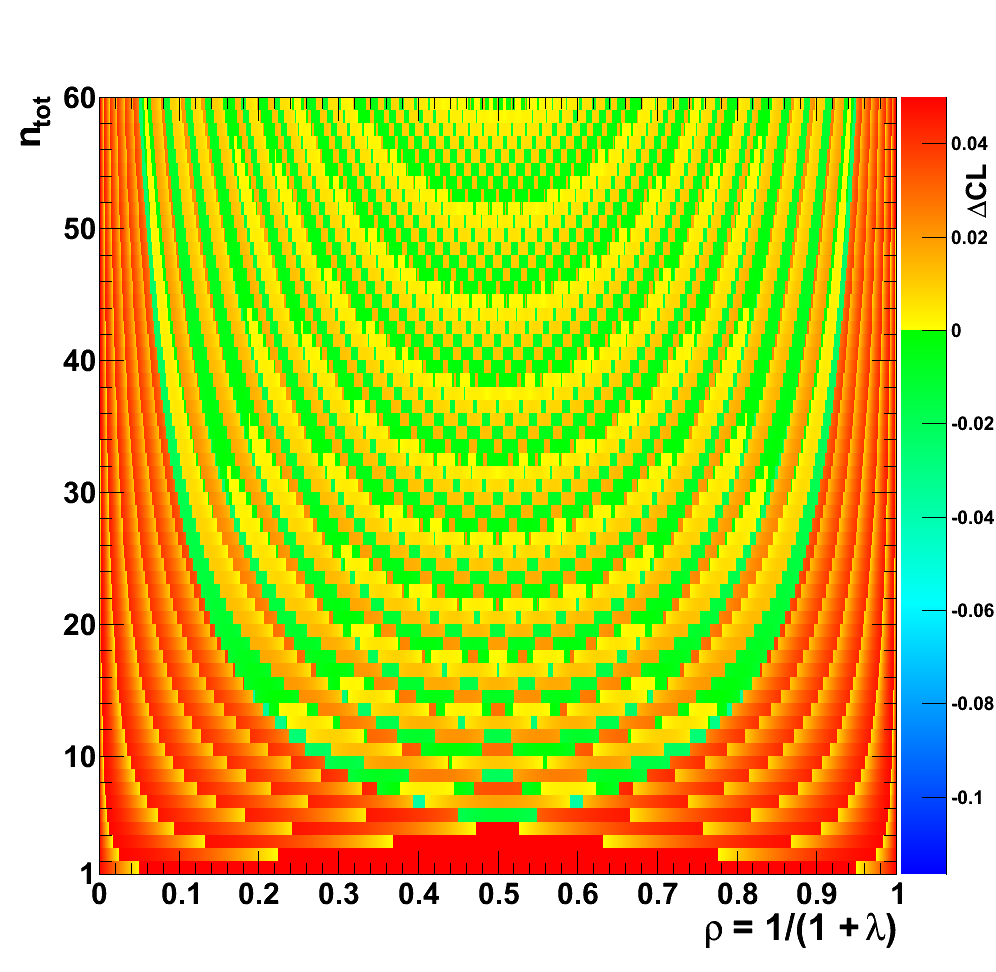}
  \put(-175,0){\bf\large (a)}
  \includegraphics*[width=0.5\textwidth]{\epsdir/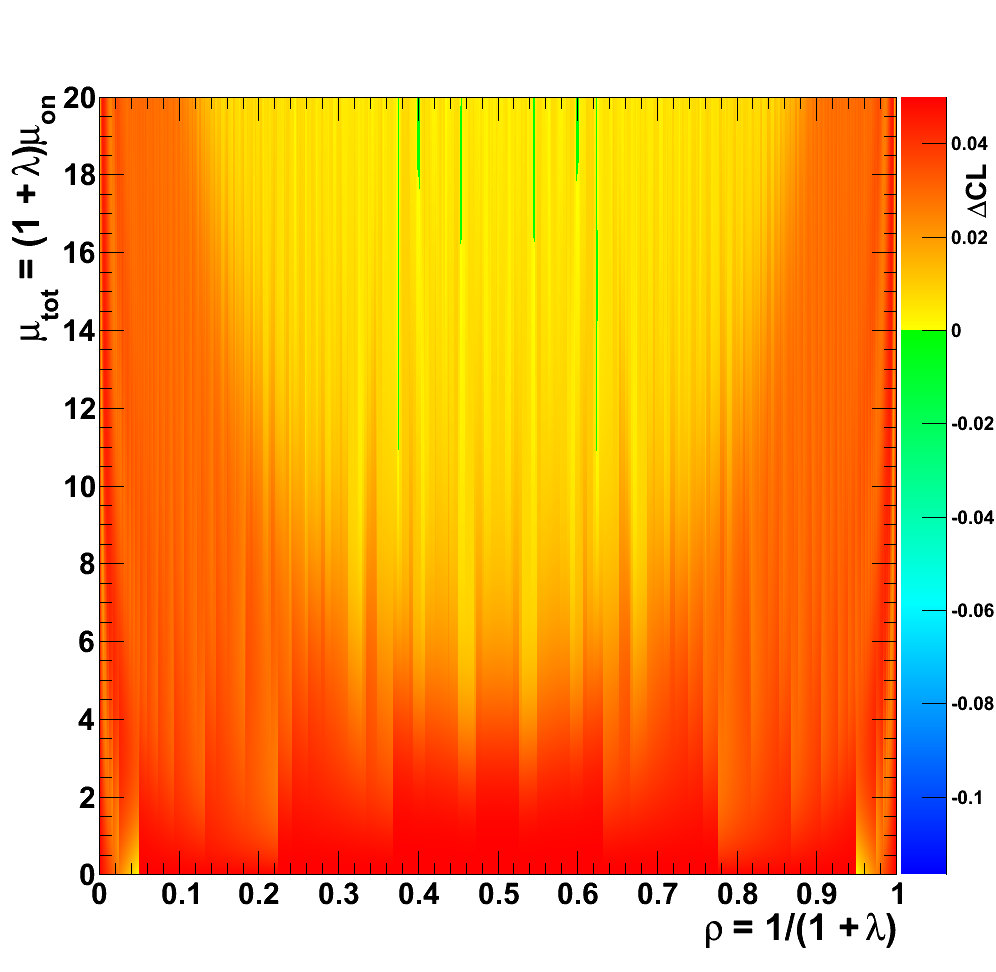}
  \put(-175,0){\bf\large (b)}
  \caption{
(a) Coverage of 95\% C.L. (Clopper-Pearson) mid-$P$ intervals, as a
function of $\binp$ and $\ntot$, and (b) unconditional coverage of the
same intervals for $\ratmean$.
}
  \label{fig:cpmidp2rat95}
\end{figure}

\begin{figure}
  \centering
  \includegraphics*[width=0.5\textwidth]{\epsdir/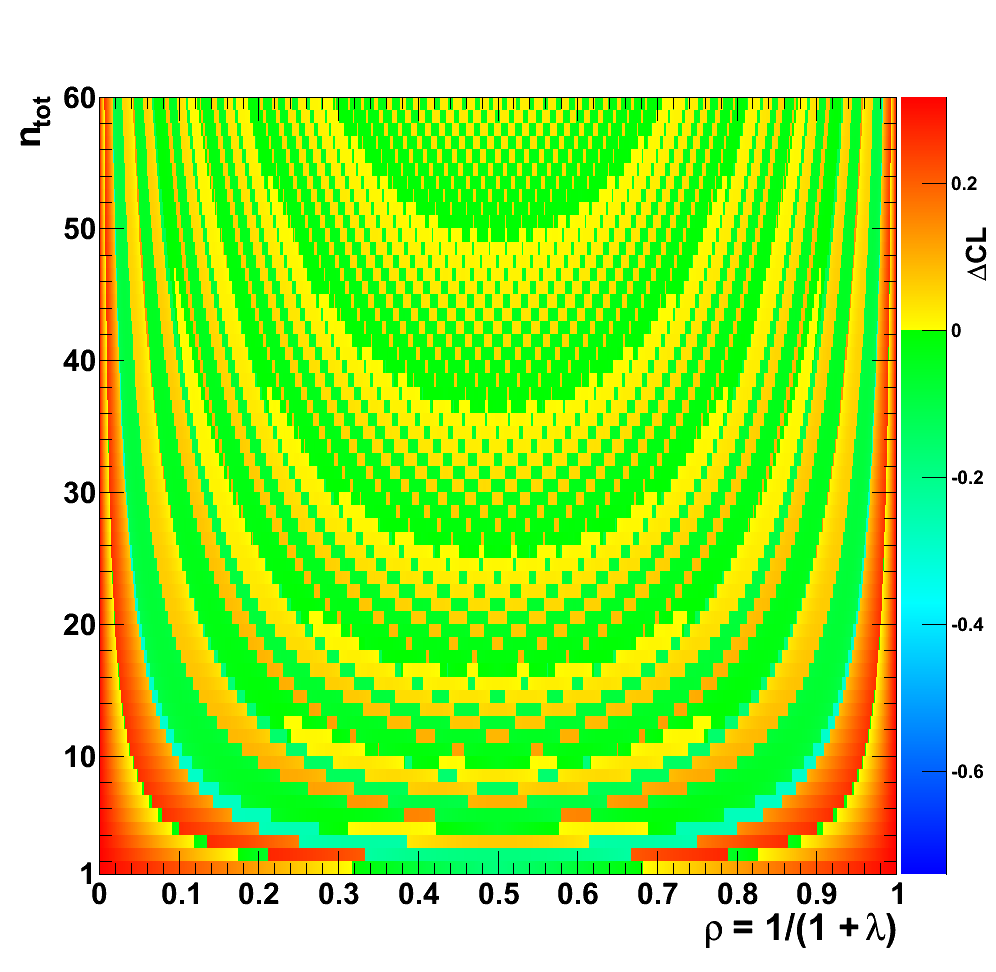}
  \put(-175,0){\bf\large (a)}
  \includegraphics*[width=0.5\textwidth]{\epsdir/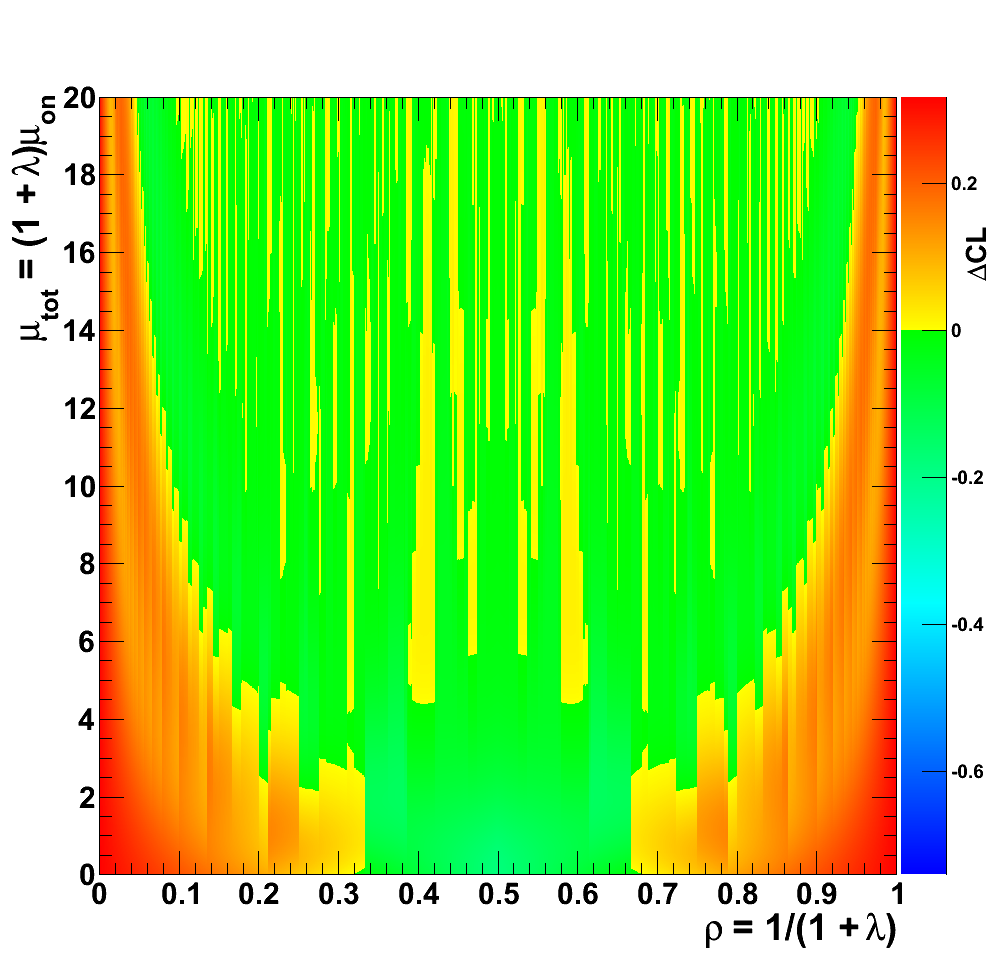}
  \put(-175,0){\bf\large (b)}
  \caption{
(a) Coverage of 68.27\% C.L. Wilson score intervals, as a function of
$\binp$ and $\ntot$, and (b) unconditional coverage of the same
intervals for $\ratmean$.
}
  \label{fig:wilson2rat}
\end{figure}

\begin{figure}
  \centering
  \includegraphics*[width=0.5\textwidth]{\epsdir/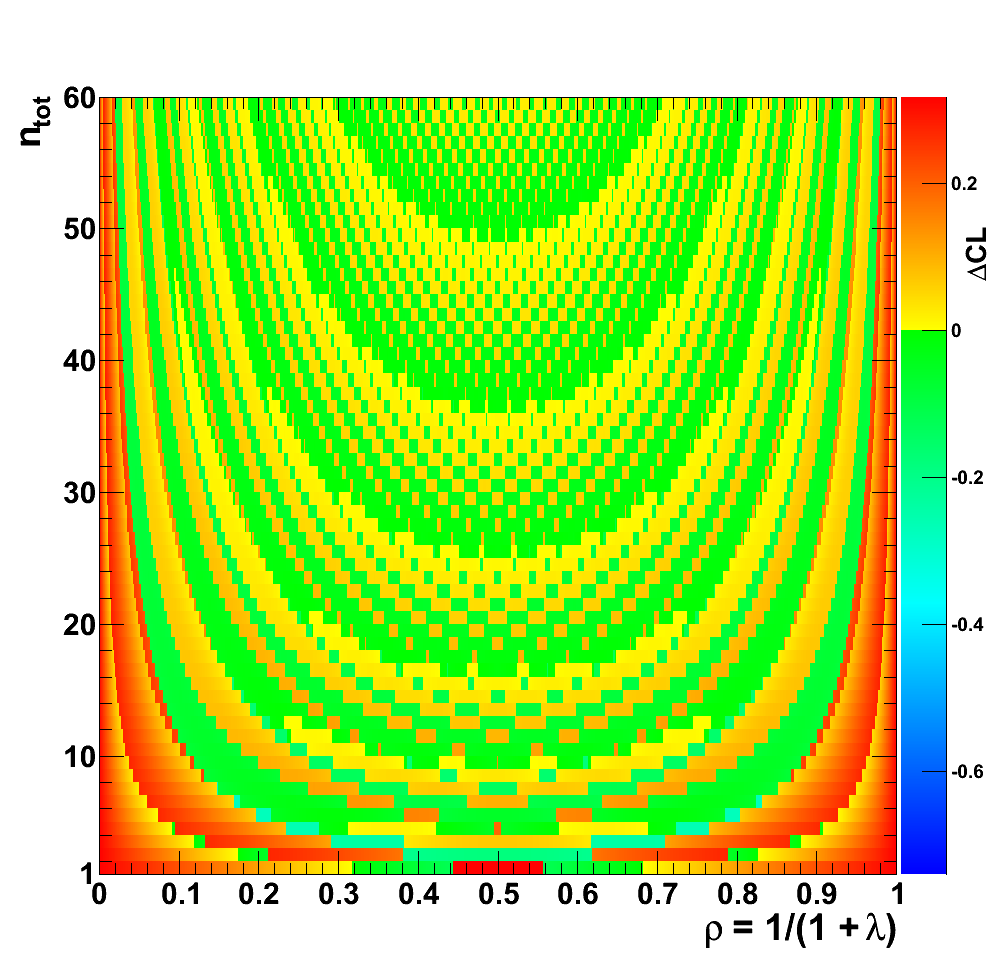}
  \put(-175,0){\bf\large (a)}
  \includegraphics*[width=0.5\textwidth]{\epsdir/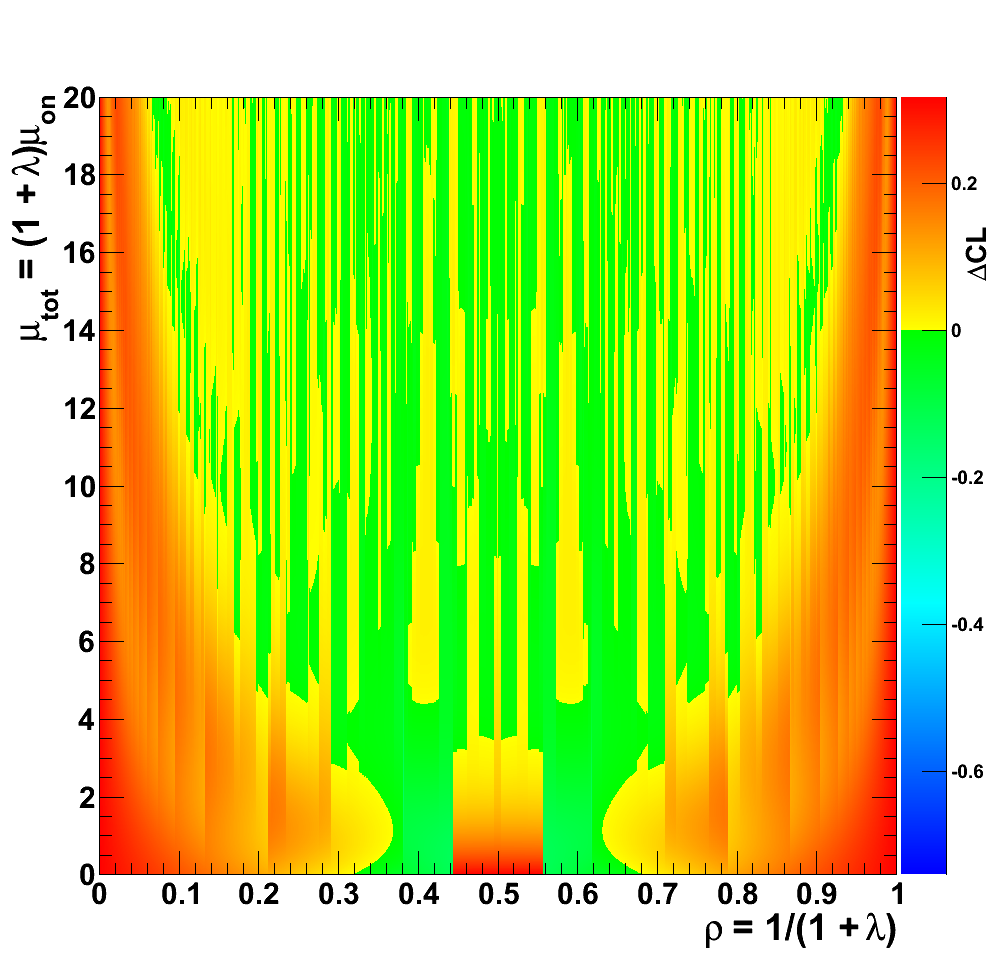}
  \put(-175,0){\bf\large (b)}
  \caption{
(a) Coverage of 68.27\% C.L. generalized Agresti-Coull intervals, as a
function of $\binp$ and $\ntot$, and (b) unconditional coverage of the
same intervals for $\ratmean$.
}
  \label{fig:ac2rat}
\end{figure}

\begin{figure}
  \centering
  \includegraphics*[width=0.5\textwidth]{\epsdir/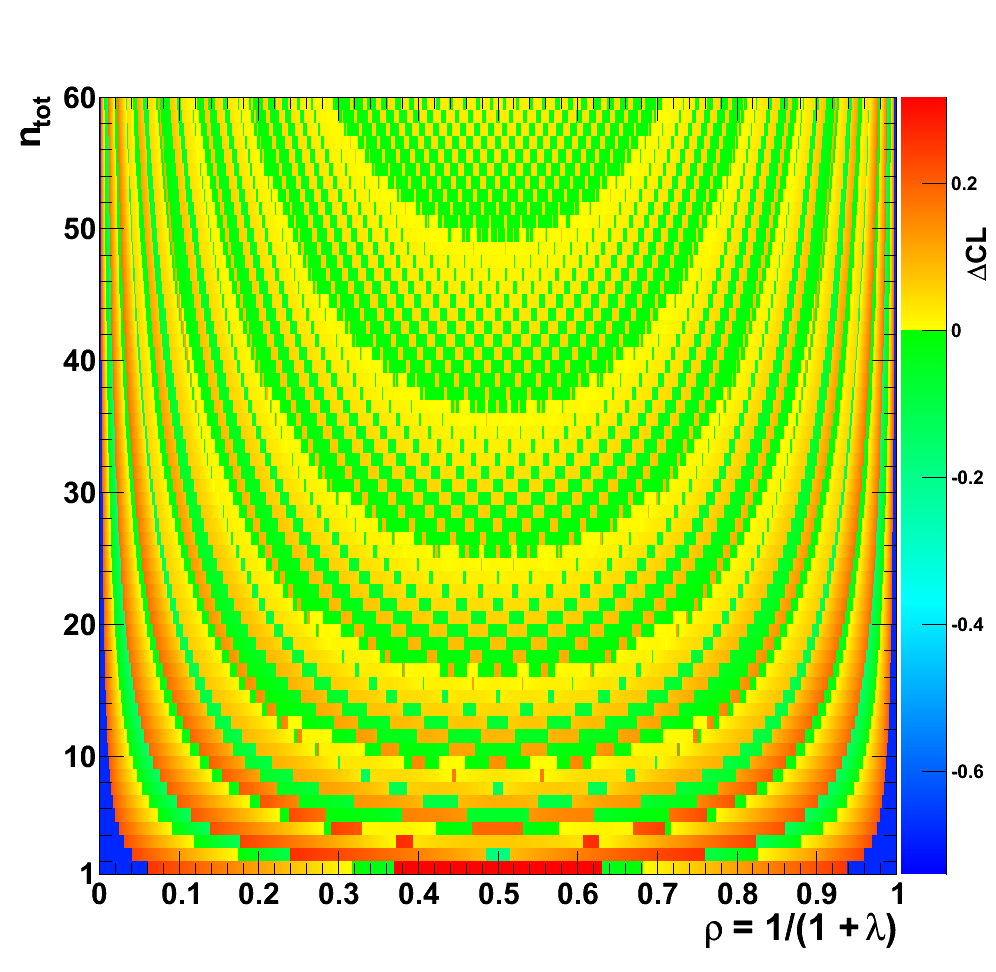}
  \put(-175,0){\bf\large (a)}
  \includegraphics*[width=0.5\textwidth]{\epsdir/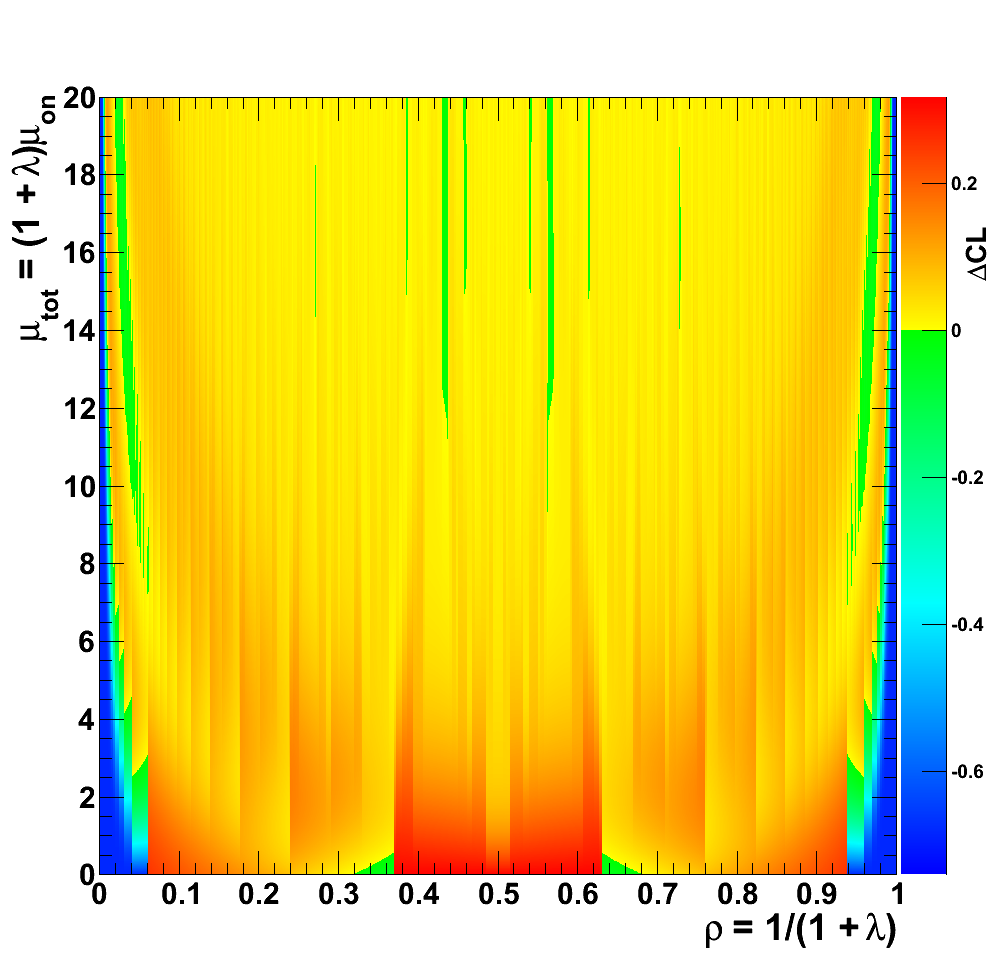}
  \put(-175,0){\bf\large (b)}
  \caption{
(a) Coverage of 68.27\% C.L. Wald-log-linear intervals, as a function
of $\binp$ and $\ntot$, and (b) unconditional coverage of the same
intervals for $\ratmean$.
}
  \label{fig:loglin2rat}
\end{figure}

\begin{figure}
  \centering
  \includegraphics*[width=0.5\textwidth]{\epsdir/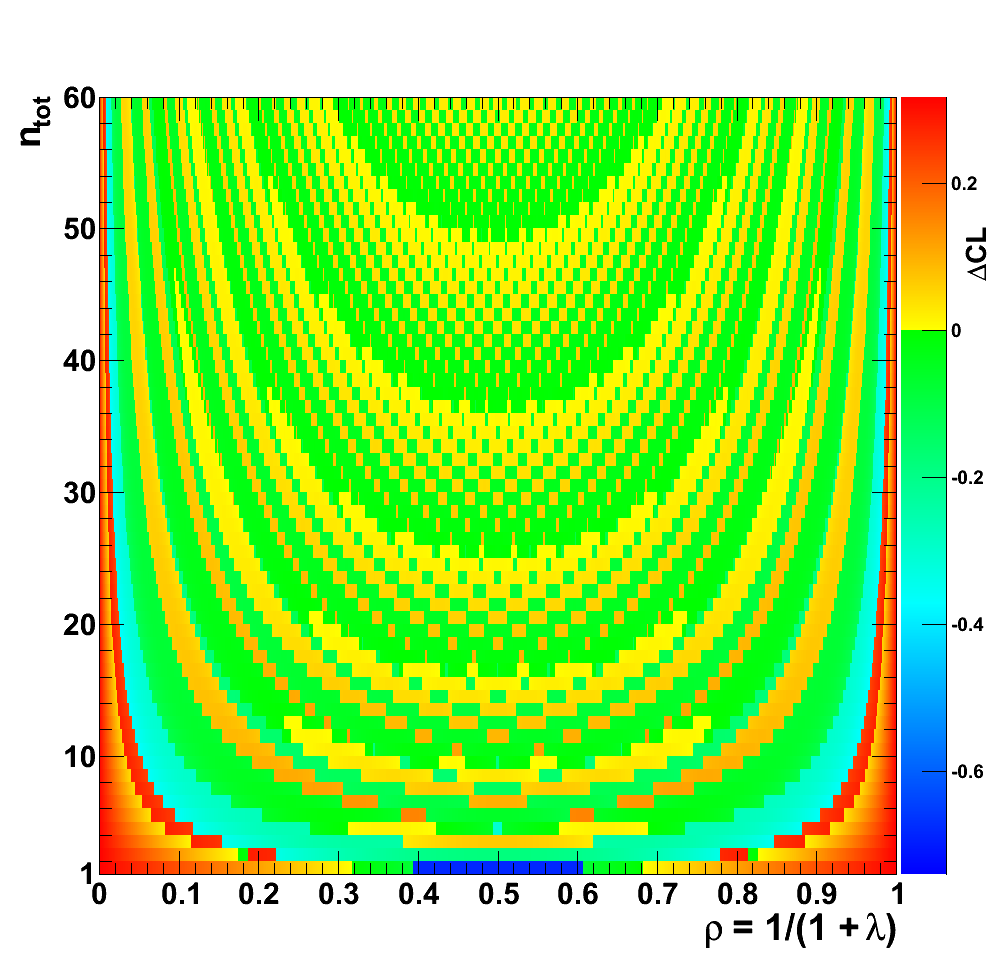}
  \put(-175,0){\bf\large (a)}
  \includegraphics*[width=0.5\textwidth]{\epsdir/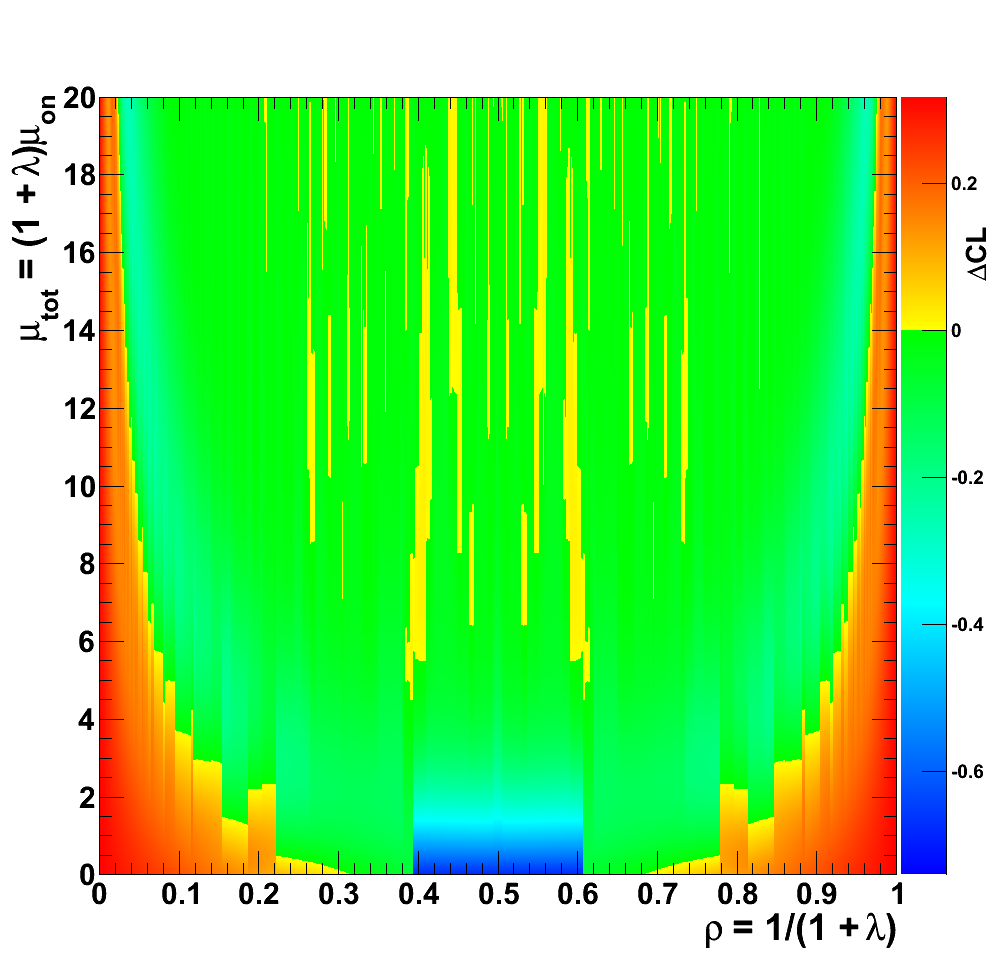}
  \put(-175,0){\bf\large (b)}
  \caption{
Coverage of 68.27\% C.L. $\deltalhood$ intervals, as a function of
$\binp$ and $\ntot$, and (b) unconditional coverage of the same
intervals for $\ratmean$.
}
  \label{fig:lhood2rat}
\end{figure}

\begin{figure}
  \centering
  \includegraphics*[width=0.5\textwidth]{\epsdir/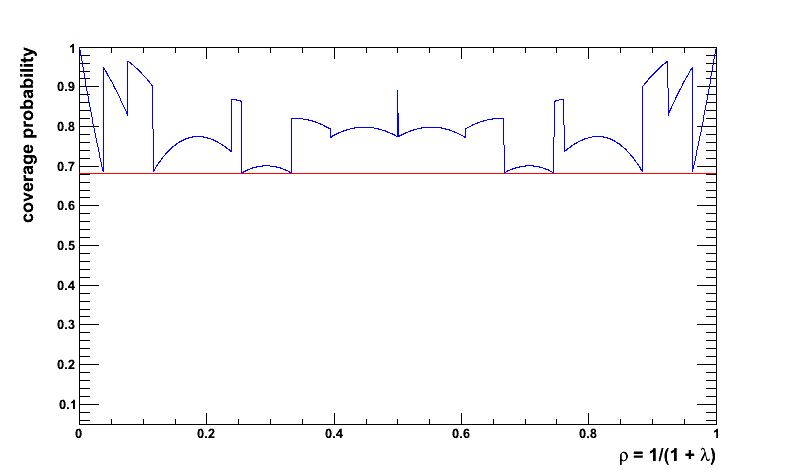}
  \put(-50,20){\bf\large (a)}
  \includegraphics*[width=0.5\textwidth]{\epsdir/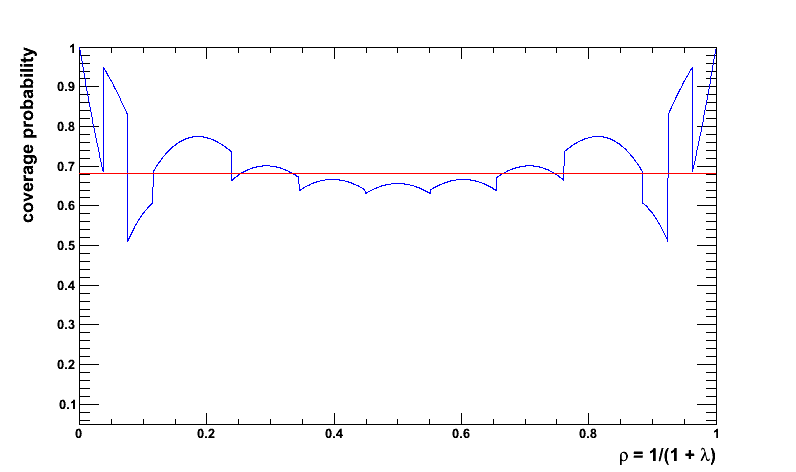}
  \put(-50,20){\bf\large (b)}
  \caption{
(a) Coverage of 68.27\% C.L. intervals obtained from exact inversion
of the LR test, as a function of $\binp$, for fixed $\ntot=10$, and
(b) coverage of same intervals but with mid-$P$ modification. (a) and
(b) are horizontal slices of Figs.~\ref{fig:fc2rat}a
and~\ref{fig:fcmidp2rat}a, respectively.
}
  \label{fig:fcrhoandmidp}
\end{figure}

\begin{figure}
  \centering
  \includegraphics*[width=0.5\textwidth]{\epsdir/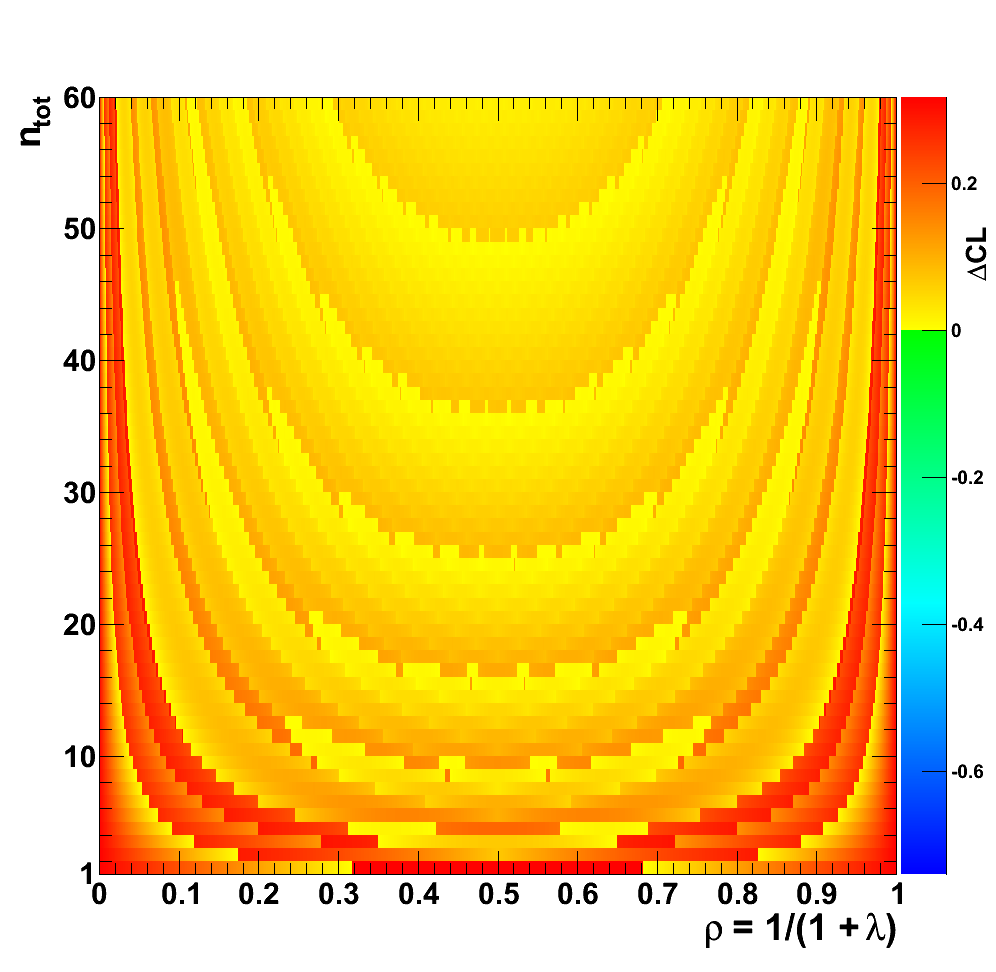}
  \put(-175,0){\bf\large (a)}
  \includegraphics*[width=0.5\textwidth]{\epsdir/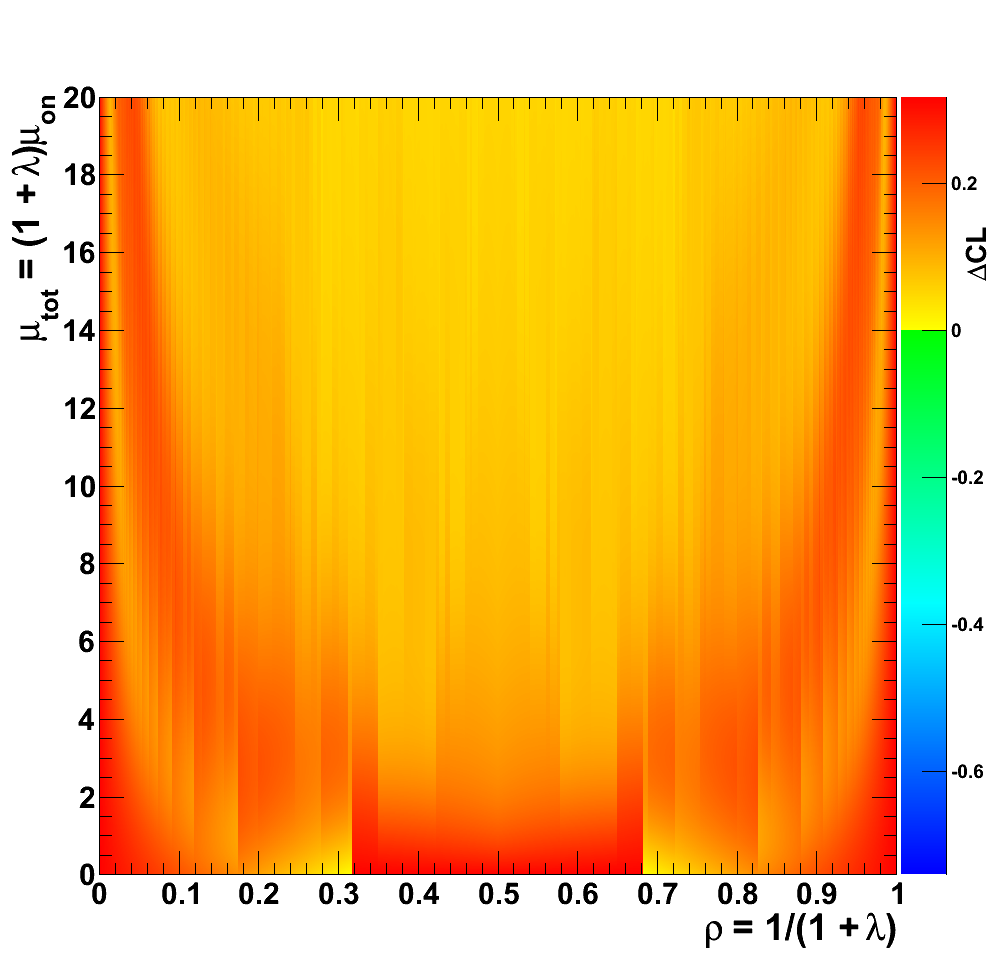}
  \put(-175,0){\bf\large (b)}
  \caption{
(a) Coverage of 68.27\% C.L. intervals obtained from exact inversion
of the LR test, as a function of $\binp$ and $\ntot$, and (b)
unconditional coverage of the same intervals for $\ratmean$. A
horizontal slice of (a) is in Fig.~\ref{fig:fcrhoandmidp}a.
}
  \label{fig:fc2rat}
\end{figure}

\begin{figure}
  \centering
  \includegraphics*[width=0.5\textwidth]{\epsdir/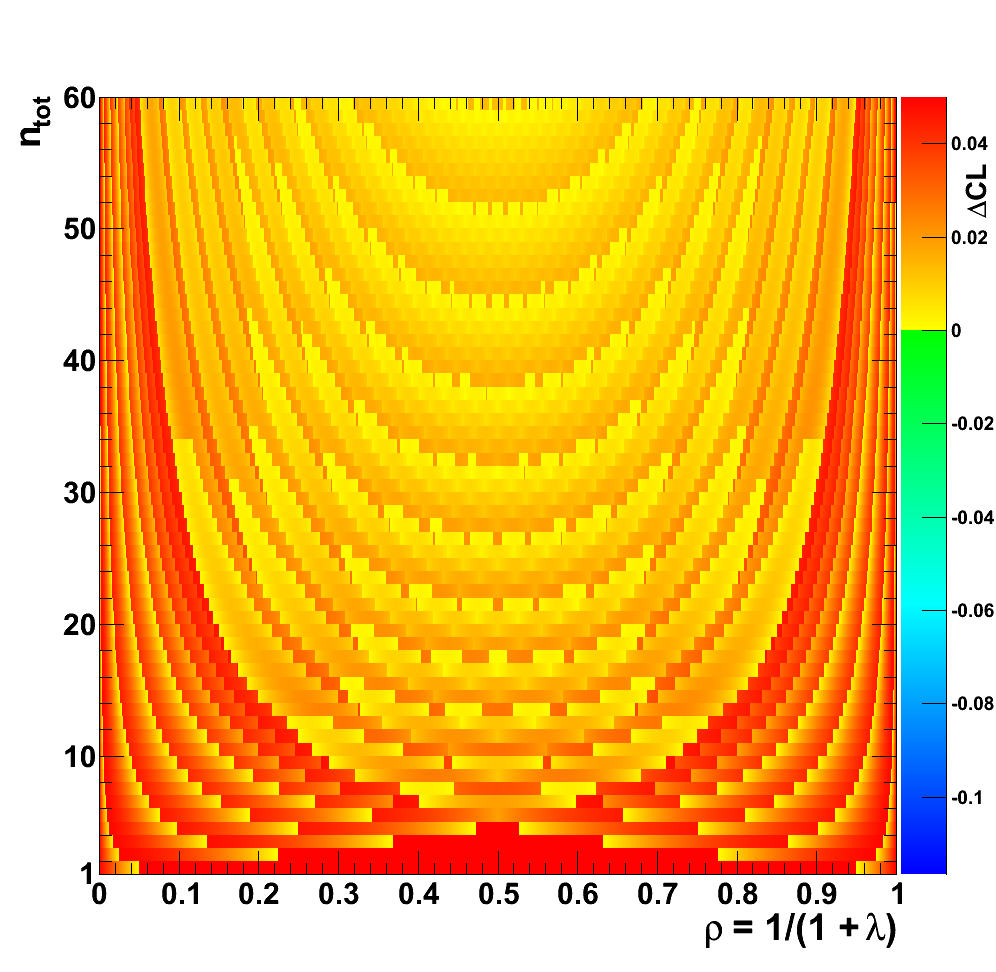}
  \put(-175,0){\bf\large (a)}
  \includegraphics*[width=0.5\textwidth]{\epsdir/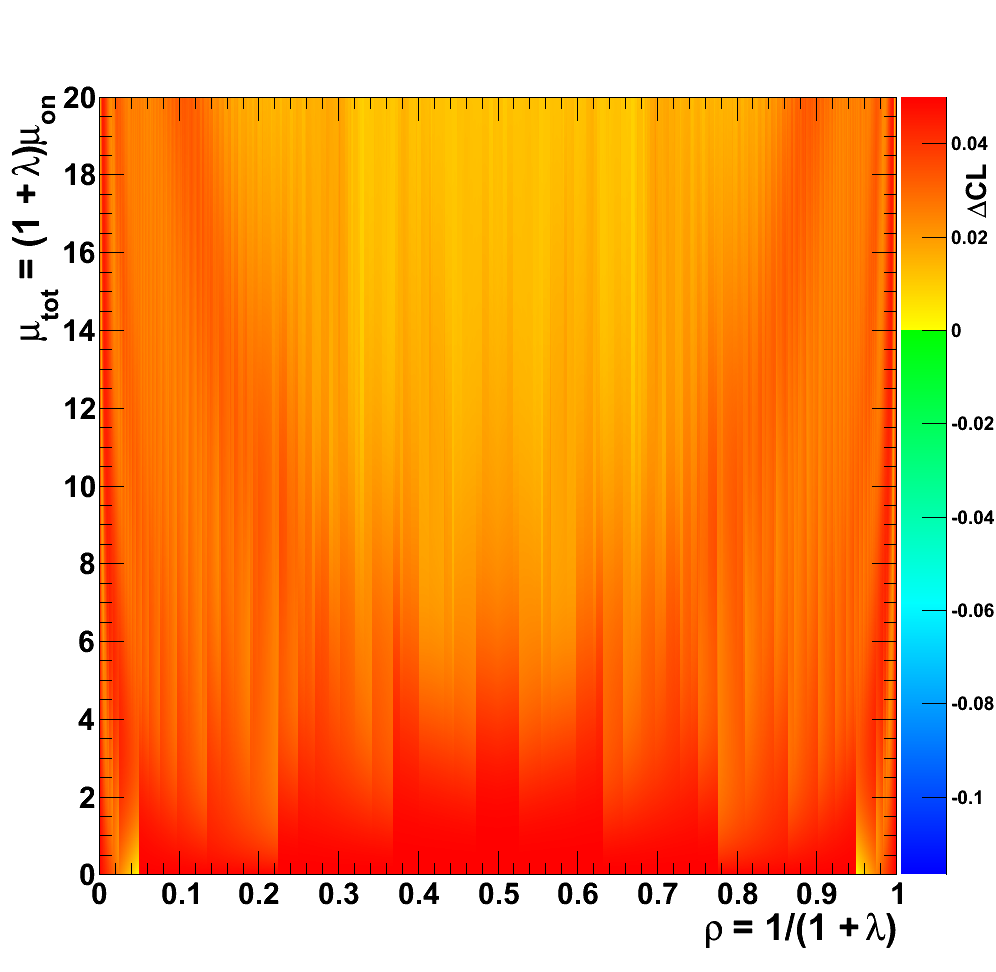}
  \put(-175,0){\bf\large (b)}
  \caption{
(a) Coverage of 95\% C.L. intervals obtained from exact inversion of
the LR test, as a function of $\binp$ and $\ntot$, and (b)
unconditional coverage of the same intervals for $\ratmean$.
}
  \label{fig:fc2rat95}
\end{figure}

\begin{figure}
  \centering
  \includegraphics*[width=0.5\textwidth]{\epsdir/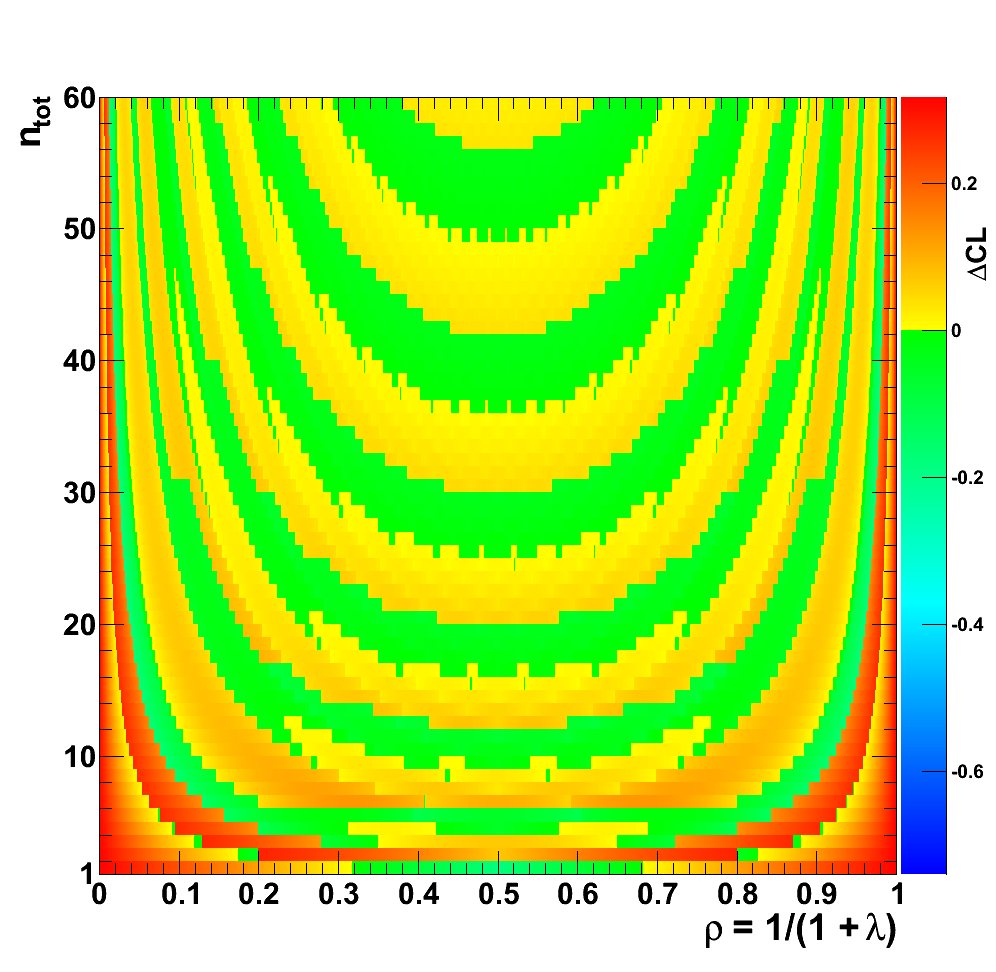}
  \put(-175,0){\bf\large (a)}
  \includegraphics*[width=0.5\textwidth]{\epsdir/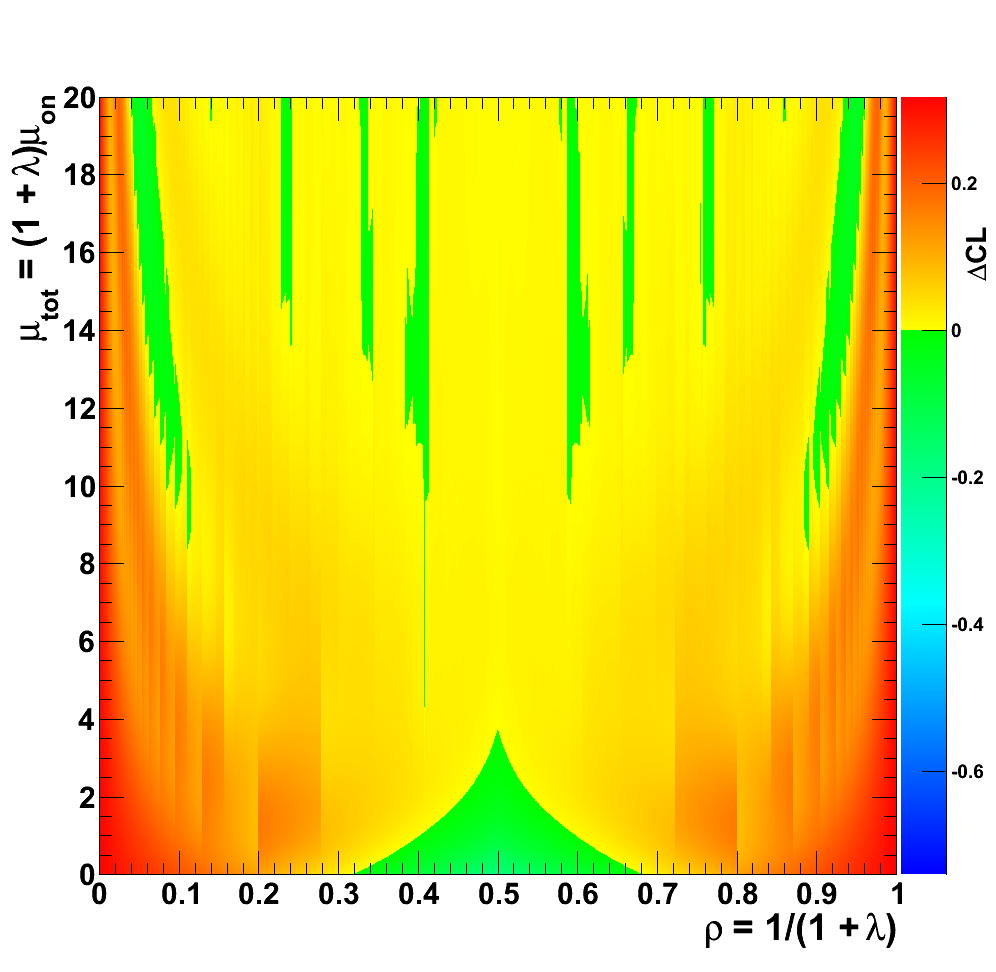}
  \put(-175,0){\bf\large (b)}
  \caption{
(a) Coverage of 68.27\% C.L. intervals obtained from exact inversion
of the LR test with mid-$P$ modification, as a function of $\binp$ and
$\ntot$, and (b) unconditional coverage of the same intervals for
$\ratmean$.  A horizontal slice of (a) is in
Fig.~\ref{fig:fcrhoandmidp}b.
}
  \label{fig:fcmidp2rat}
\end{figure}

\begin{figure}
  \centering
  \includegraphics*[width=0.5\textwidth]{\epsdir/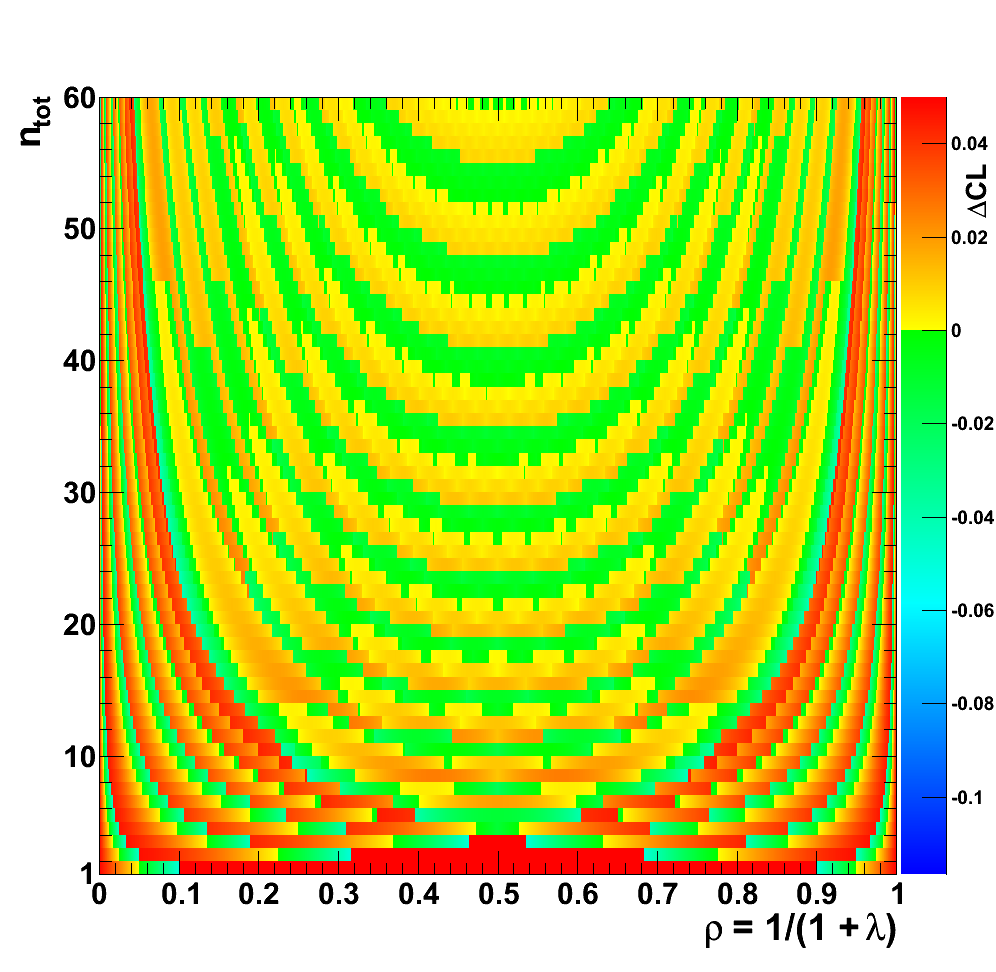}
  \put(-175,0){\bf\large (a)}
  \includegraphics*[width=0.5\textwidth]{\epsdir/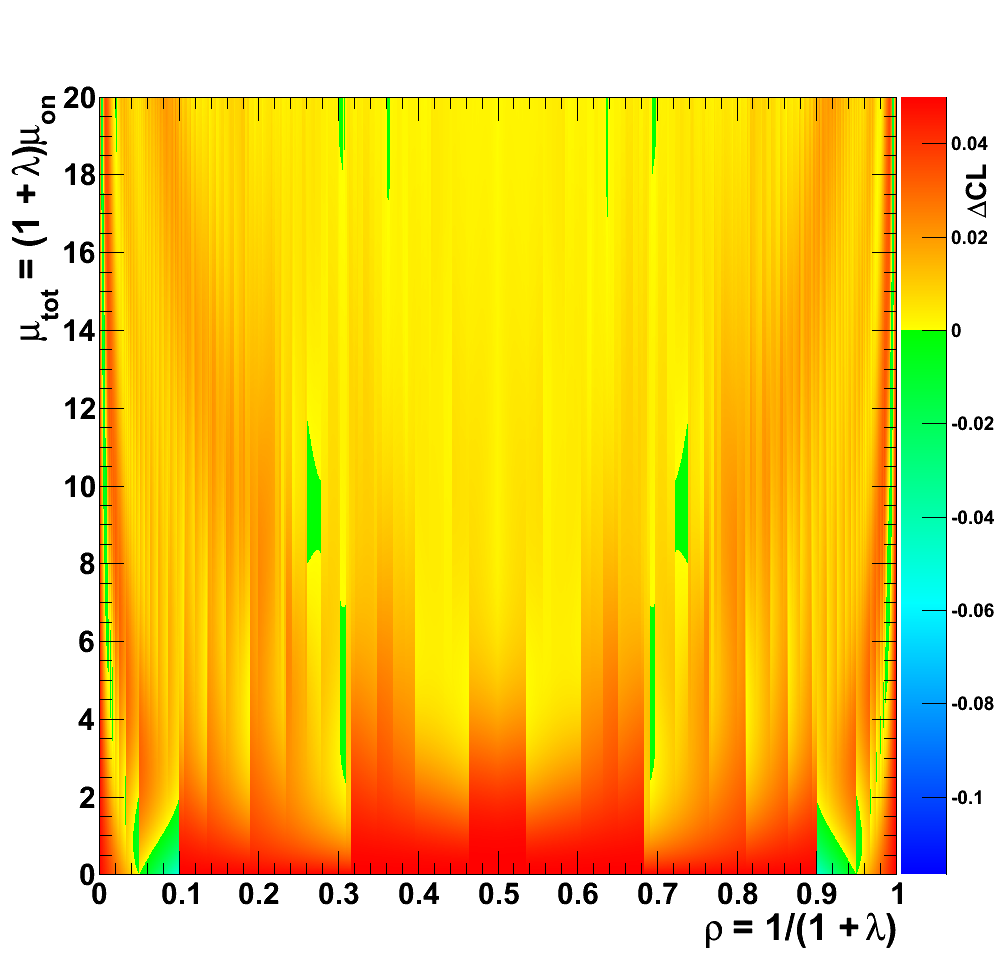}
  \put(-175,0){\bf\large (b)}
  \caption{
(a) Coverage of 95\% C.L. intervals obtained from exact inversion of
the LR test with mid-$P$ modification, as a function of $\binp$ and
$\ntot$, and (b) unconditional coverage of the same intervals for
$\ratmean$.
}
  \label{fig:fcmidp2rat95}
\end{figure}

\begin{figure}
  \centering
  \includegraphics*[width=0.5\textwidth]{\epsdir/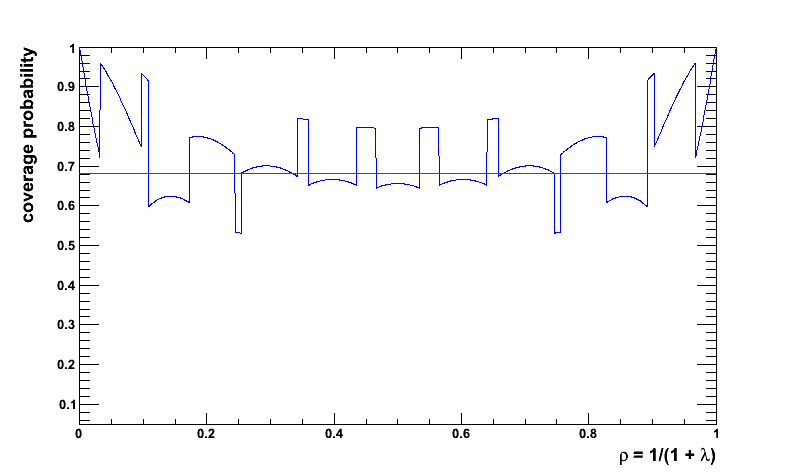}
  \put(-50,20){\bf\large (a)}
  \includegraphics*[width=0.5\textwidth]{\epsdir/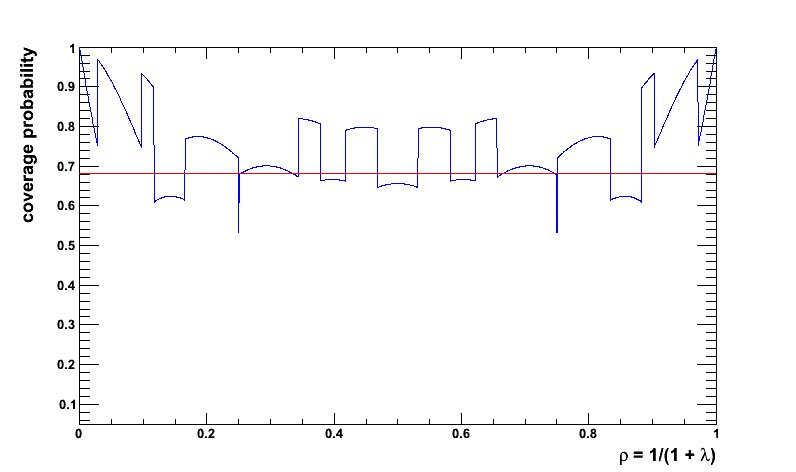}
  \put(-50,20){\bf\large (b)}
  \caption{
(a) Coverage of 68.27\% C.L. (Clopper-Pearson) mid-$P$ intervals, and
(b) coverage of 68.27\% C.L. intervals constructed by
Cousins \cite{cousinsratio} for the ratio of Poisson means and
translated here to intervals for $\binp$, both as a function of
$\binp$, for fixed $\ntot=10$. (a) and (b) are horizontal slices of
Figs.~\ref{fig:cpmidp2rat}a and~\ref{fig:cousins2rat}a, respectively.
The remarkable resemblance is typical
of that for other values of $\binp$ and $\ntot$.}
  \label{fig:cpmidpcousinsrho}
\end{figure}

\begin{figure}
  \centering
  \includegraphics*[width=0.5\textwidth]{\epsdir/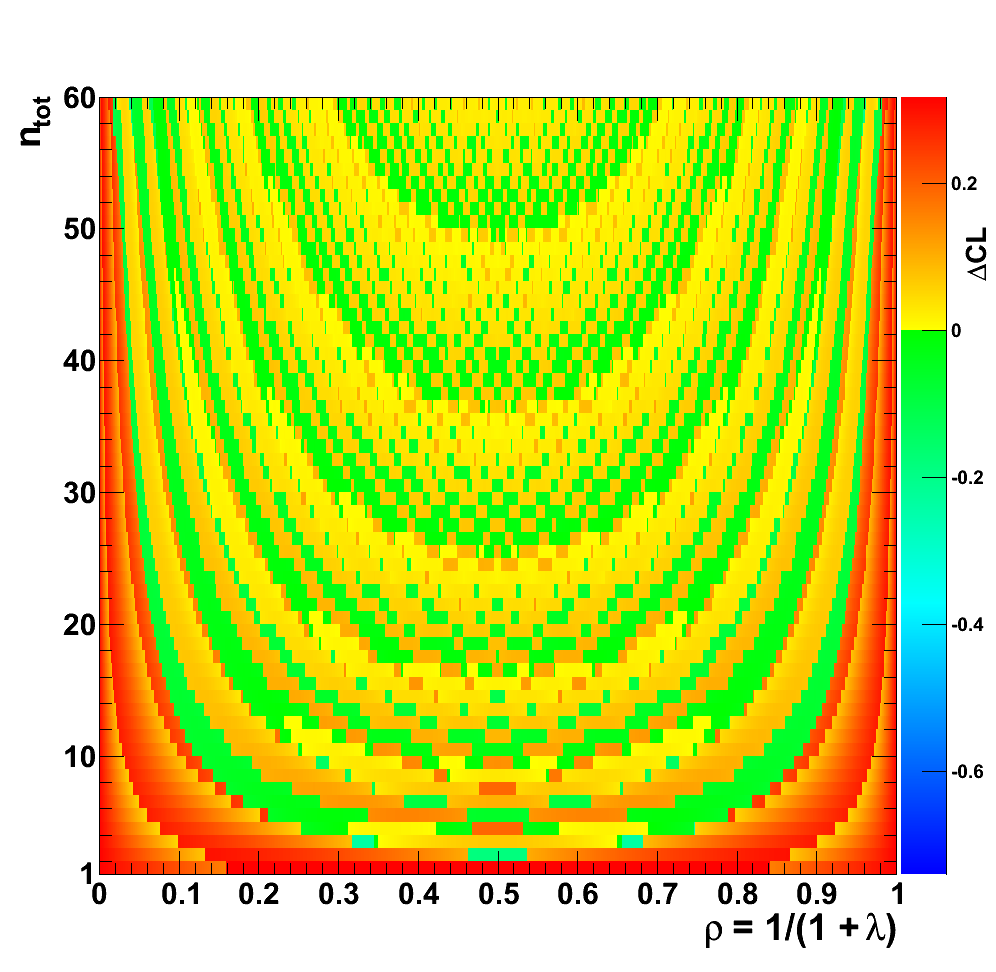}
  \put(-175,0){\bf\large (a)}
  \includegraphics*[width=0.5\textwidth]{\epsdir/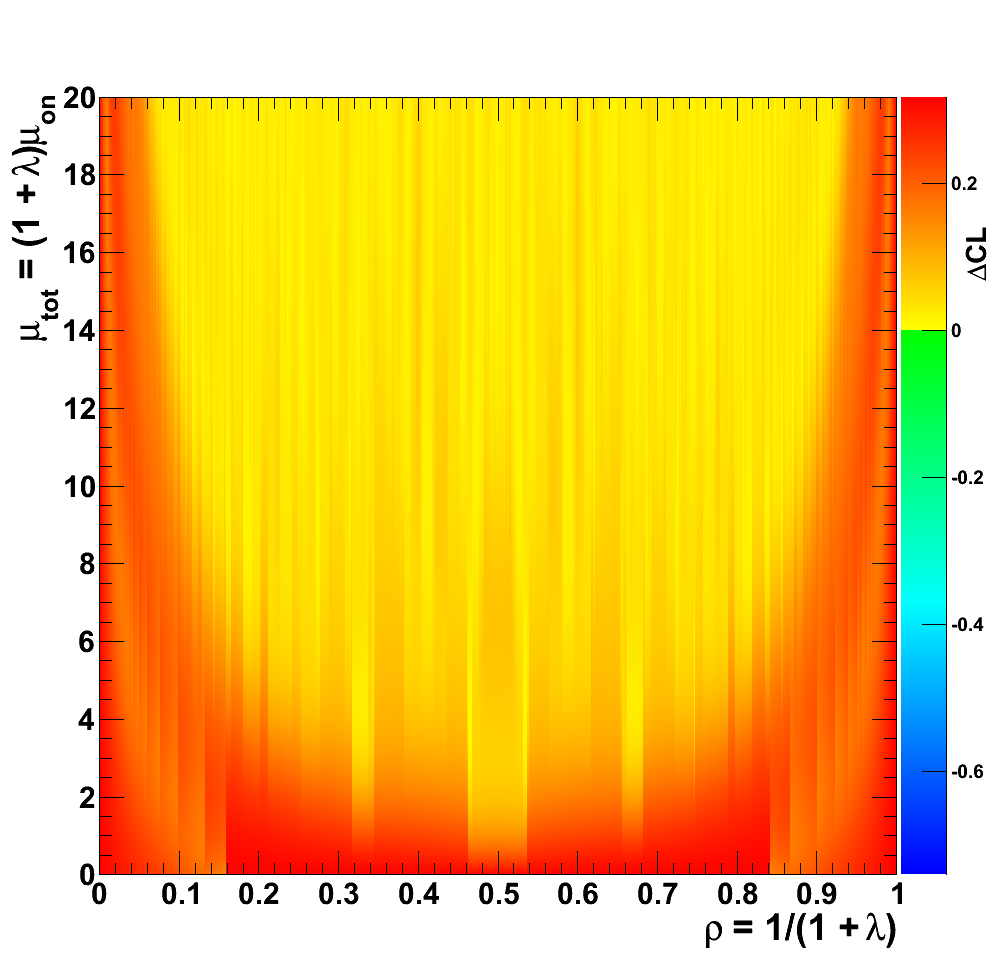}
  \put(-175,0){\bf\large (b)}
  \caption{
(a) Coverage of 68.27\% C.L. intervals constructed by Cousins
\cite{cousinsratio} for the ratio of Poisson means and translated here
to intervals for $\binp$, as a function of $\binp$ and $\ntot$, and
(b) unconditional coverage of the same intervals for $\ratmean$. A
horizontal slice of (a) is in Fig.~\ref{fig:cpmidpcousinsrho}b.
}
  \label{fig:cousins2rat}
\end{figure}

\begin{figure}
  \centering
  \includegraphics*[width=0.5\textwidth]{\epsdir/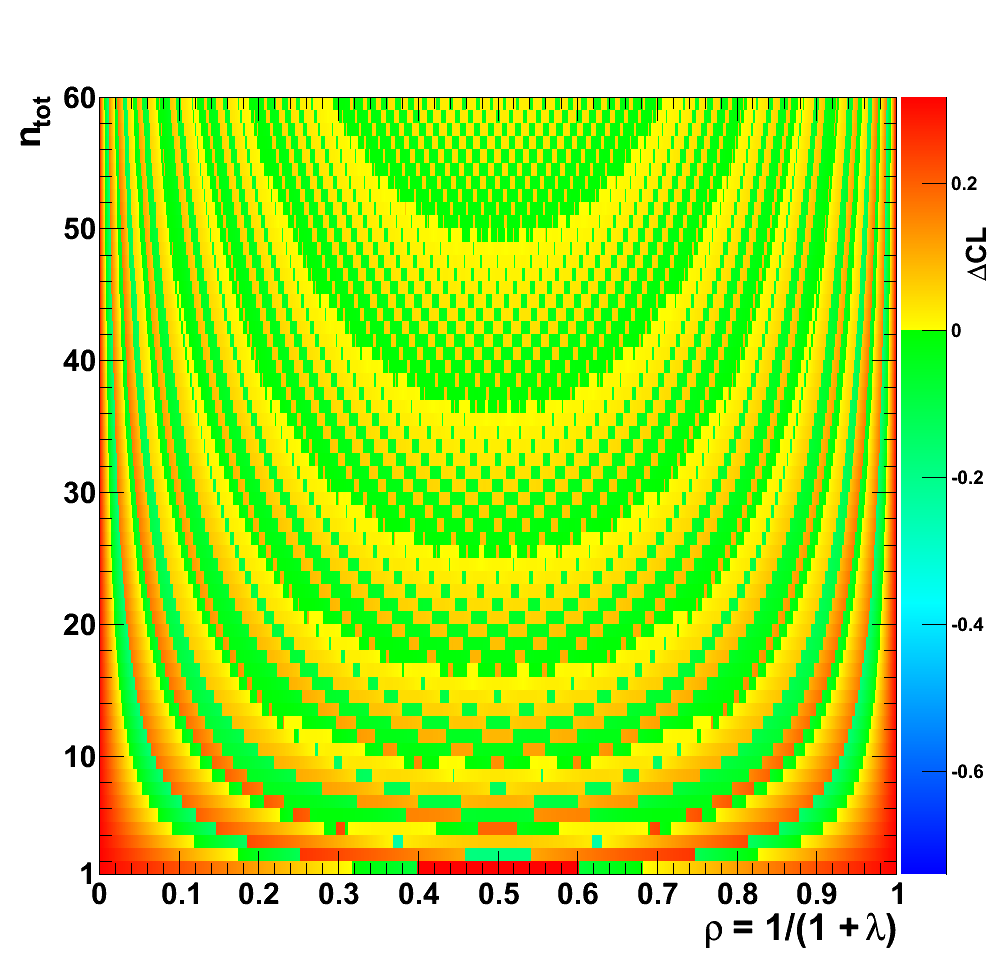}
  \put(-175,0){\bf\large (a)}
  \includegraphics*[width=0.5\textwidth]{\epsdir/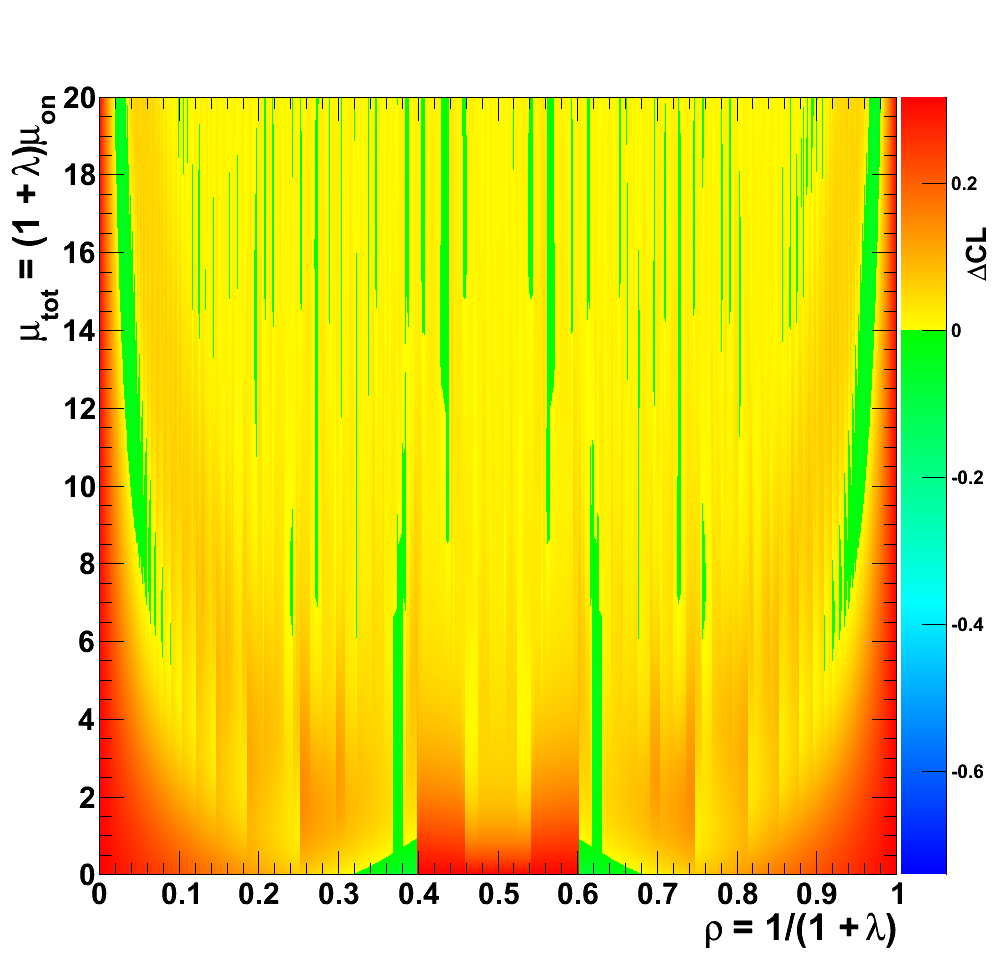}
  \put(-175,0){\bf\large (b)}
  \caption{
(a) Coverage of intervals calculated using a Bayesian method with
uniform prior and containing 68.27\% posterior probability, as a
function of $\binp$ and $\ntot$, and (b) unconditional coverage of the
same intervals for $\ratmean$.
}
  \label{fig:uniform2rat}
\end{figure}

%C-P and Jeffreys are already above
In Figs.~\ref{fig:cpmidp2rat} through \ref{fig:uniform2rat}, we plot the
probability that the parameter is in the interval for methods which
are among those advocated in the above references: Clopper-Pearson
with mid-$P$ modification, at 68.27\% C.L. (Figs.~\ref{fig:cpmidp2rat}a and b, \ref{fig:cpmidpcousinsrho}a)
and at 95\%
C.L. (Figs.~\ref{fig:cpmidp2rat95}a and b); Wilson score
at 68.27\% C.L. (Figs.~\ref{fig:wilson2rat}a and b);
generalized Agresti-Coull at 68.27\% C.L. (Figs.~\ref{fig:ac2rat}a and b);
Wald log-linear at 68.27\%
C.L. (Figs.~\ref{fig:loglin2rat}a and b); $\deltalhood$ at
68.27\% C.L. (Figs.~\ref{fig:lhood2rat}a and b); exact LR
test inversion, i.e., Neyman construction with likelihood-ratio
ordering, with and without mid-$P$ modification, at both C.L.'s
(Figs.~\ref{fig:fcrhoandmidp} through \ref{fig:fcmidp2rat95}); Cousins's
\cite{cousinsratio} ratio-of-Poisson means translated into binomial at
68.27\% C.L.  (Figs.~\ref{fig:cpmidpcousinsrho}b, \ref{fig:cousins2rat}a and b);
and Bayesian with prior uniform in $\binp$ at
68.27\% C.L. (Figs.~\ref{fig:uniform2rat}a and b).
Additional plots for nearly all methods mentioned in this paper, for a
variety of confidence levels, slices, as well as for one-sided 
probabilities, are available on request from the authors.

A striking aspect of the two-dimensional plots is the variation of
coverage, which is difficult to capture in tables of average coverage
or rms of coverage: superimposed on the oscillations are evident
trends which indicate regions of particularly low or high coverage.  A
number of methods give large undercoverage either at low $\ntot$ or at
$\binp$ near the endpoints; as these values naturally arise in HEP
applications, we disfavor such methods.

Another significant observation is how well the mid-$P$ methods
perform in the unconditional coverage calculations for the ratio of
Poisson means.  In effect the Poisson fluctuations of $\ntot$ are
performing some randomization on top of the ``mid'' value (0.5) which
was fixed in the mid-$P$ calculation for fixed $\ntot$.  For central
intervals, the result is a remarkable resemblance to the corresponding
plots for the central intervals of Cousins \cite{cousinsratio}, which
are strictly conservative for the ratio of Poisson means, but much
less so than Clopper-Pearson intervals. This similarity was discovered
while performing the calculations for this paper, and is seen in 
Figs.~\ref{fig:cpmidpcousinsrho}a and b;
Figs.~\ref{fig:cpmidp2rat}a and~\ref{fig:cousins2rat}a;
Figs.~\ref{fig:cpmidp2rat}b and~\ref{fig:cousins2rat}b; 
and in numerous other plots 
inspected by the authors. One could imagine
further tuning (as a function of $\ntot$) the ``mid'' value of 0.5 in
order to optimize coverage, but we did not explore this.

The Bayesian-inspired methods perform reasonably well with the ad-hoc
choice of using central intervals unless $\non$ is 0 or $\ntot$, in
which case the interval was pushed against the endpoint, as described
above.  This leads to over-coverage near the endpoints (a feature of
many methods).  We did not explore alternatives such as
highest-posterior-density intervals.

Among asymptotic methods, the Wilson score interval and the
(preferred) generalized Agresti-Coull interval appear to be reasonable
for quick estimates as various authors have advocated.  The
$\deltalhood$ method undercovers at low $\ntot$, and is generally not
advocated in the literature reviewed.

\section{Conclusion}
\label{conclusion}

While intervals such as the Wilson score and the generalized
Agresti-Coull can be useful for hand calculations and quick estimates
(and are a dramatic improvement over the Wald intervals), the methods
based on ``exact'' calculations (i.e., using the binomial and Poisson
probabilities rather than asymptotic or Bayesian-inspired
calculations) appear to give the most reliable frequentist coverage.
When strictly conservative coverage is desired, this statement is a
tautology, but it also appears to be the case when approximate
coverage is desired, if (as we advocate) the average coverage is
evaluated by averaging over data in the closely related
ratio-of-Poisson-means problem, rather than attempting to average over
$\binp$. 

For {\it central} intervals, the original {\it
Clopper-Pearson} intervals \cite{clopper34} remain the strictly
conservative standard \cite{pdg}, 
at the cost of severe over-coverage, especially
at small $\ntot$.  Among the many variants of strictly conservative
{\it both-tailed} (non-central) intervals, we prefer those based on
{\it likelihood-ratio-ordering}, i.e., the intervals obtained by
``exact inversion of the LR test'', the method advocated in HEP by Feldman
and Cousins \cite{feldmancousins}. The
likelihood-ratio test \cite{kendall} generalizes well to many complex,
multi-dimensional problems in statistical inference
\cite{feldmancousins}, and thus is well-integrated into a larger
picture; when using more specialized ad hoc manipulations applied to
the binomial problem, one is faced with the problem of when to abandon
them (and what to replace them with) as more complications are added 
to the original simple $(\non,\ntot)$ problem.

In the ratio-of-Poisson-means problem, we prefer making {\it
Lancaster's mid-$P$ modification} \cite{lancaster61} to the
construction of either set of exact intervals in the above paragraph.
It provides remarkably good approximate coverage in the ratio-of-Poisson-means
problem when evaluated in the unconditional ensemble (i.e.,
frequentist averaging over values of $\ntot$ other than the value
observed, weighted by their Poisson probabilities). The mid-$P$
intervals are strikingly similar to a set constructed by Cousins which
strictly covers the ratio, but the mid-$P$ intervals have a much
simpler description that can also be generalized as complexity is
added to the problem.  One can also imagine contexts (such as
estimating efficiencies of many similar detector elements) in which
Poisson fluctuations of the number of trials in each detector element
provides a sort of frequentist ensemble which would suggest that
mid-$P$ intervals should be considered.
However, use of mid-$P$ intervals in a context
in which there is no such frequentist averaging would go against the
traditional conventions of HEP.  Introduction of nuisance parameters
(e.g., some systematic uncertainties) into the pure binomial problem,
as is common in HEP, can provide another source of averaging.  We
speculate that mid-$P$ intervals could prove to be useful for obtaining
good coverage in many such contexts.

The use of these intervals can of course be considered in any
application of binomial intervals.  In high energy and astroparticle
physics, the ``on/off'' (signal bin plus sideband) problem was
recently explored in detail by Cousins, Linnemann, and Tucker
\cite{clt}; one of the promising methods for computing the statistical
significance of a signal (denoted by $\zbi$) used the Clopper-Pearson
interval.  In some contexts it should be useful to consider as well
one or more of the other three intervals recommended here when
calculating $\zbi$.

% \ack is defined in elsart.cls to be the acknowledgement command!
\ack{This work was partially supported by the U.S. Department of
Energy and the National Science Foundation.}

\clearpage

% \appendix
%\section{Notation}
%\label{app:notation}

\end{document}